\documentclass[%
aip,
amsmath,amssymb,
reprint,%
]{revtex4-1}

\usepackage{graphicx}
\usepackage{dcolumn}
\usepackage{bm}

\usepackage[utf8]{inputenc}
\usepackage[T1]{fontenc}
\usepackage{mathptmx}
\usepackage{etoolbox}
\usepackage{graphicx}
\usepackage{capt-of}

\makeatletter
\def\@email#1#2{%
	\endgroup
	\patchcmd{\titleblock@produce}
	{\frontmatter@RRAPformat}
	{\frontmatter@RRAPformat{\produce@RRAP{*#1\href{mailto:#2}{#2}}}\frontmatter@RRAPformat}
	{}{}
}%
\makeatother
\begin{document}
	
	\preprint{AIP/123-QED}
	
\title[]{Vibration patterns and reed–pipe coupling in the free reeds of the \emph{hulusi}}
\author{Yang Xia}
\affiliation{College of Art, Zhejiang Normal University, Jinhua 321004, China}
\affiliation{Institute of Systematic Musicology, University of Hamburg, Hamburg 20354, Germany}
\author{Rolf Bader}
\affiliation{Institute of Systematic Musicology, University of Hamburg, Hamburg 20354, Germany}
\author{Simon Linke}
\affiliation{Hamburg University for Applied Sciences, Ligeti Center, Hamburg 21079, Germany}
\email{R\_Bader@t-online.de}

\date{\today} 

\begin{abstract}
The \emph{hulusi} is a free-reed instrument from Yunnan, Western
China, known for its unique sound. Its reed and frame are cut from a single
brass sheet in a symmetric shape, the free end narrower than the clamped
end. Its free reeds are coupled to tubes that determine the instrument's pitch during normal playing. With reduced playing pressure, the reed frequency takes over the instrument sound.  We report
point-by-point laser Doppler vibrometry on the reeds of six handmade
\emph{hulusi}, two each in the keys of F, B$\flat$ and C, measured at 29
points on every reed or tongue over nine fingerings and two supply pressures.
Steady-state spectra carry two components: a tube-dominant frequency
$f_\mathrm{tube}$, locked to the air-column resonance and carrying a
complete harmonic series, and a reed-dominant frequency
$f_\mathrm{reed}$, which is not an integer multiple of it. Their
difference $\Delta = f_\mathrm{tube} - f_\mathrm{reed}$ runs from 38 to
523~Hz across the dataset, from about three semitones to more than an
octave. The reed-dominant frequency rises
with the pressure at the reed, by $28.0 \pm 10.3$~Hz/kPa once
$f_\mathrm{tube}$ is controlled for. At low pressure supply the oscillation loses
its lock to the air column, falls silent for about 100~ms, and returns
at the reed frequency within 0.1~Hz. The spatial patterns are
forced oscillation shapes of one limit cycle at its fundamental and
harmonics rather than eigenmodes: the measured frequencies stand
in strict 1:2:3 ratios, far from the bending ratios of a cantilever. The
reed-dominant pattern is pure bending and holds across fingerings and
supply pressures, whereas the tube-dominant pattern breaks up near the
oscillation threshold.
\end{abstract}
\maketitle


\section{\label{sec:1} Introduction}
As a traditional Chinese free-reed instrument, the unique tonal characteristics of the \emph{hulusi} depend largely on the vibrational properties of its reed. The reed is the core sound-producing component of the \emph{hulusi}; its material properties, geometric shape, and vibration modes directly influence the instrument's timbre, intonation, and playing stability\citep{Dieckman2006}.

The instrument is shown in Fig.~1. Air enters through the mouthpiece into the gourd, which serves as the wind chamber, and three pipes are set into the lower end of the gourd. The \emph{hulusi} reed is cut from a single piece of material (usually brass) in a left-right symmetrical shape and is mounted in a window cut in the wall of the socket at the upper end of each pipe [Fig.~1(b)]. The left and right edges of the reed are thinned by hand, and the reed is then attached to the tubular resonator\citep{Yang2017Instrument}. Airflow through the gourd sets the reed vibrating. Under normal playing conditions, the instrument itself acts as the primary oscillator, while the reed locks onto the instrument's acoustic frequency\citep{Fletcher1978}. Only the middle pipe carries finger holes; different combinations of open and closed holes alter its resonant characteristics and excite standing waves at different frequencies. The two outer pipes are drones, with reeds of the same shape but no finger holes, so their pitch cannot be changed, and the measurements reported here are of the melodic pipe alone. Studies have shown that the sounding frequency of the reed--tube combination is higher than both the reed's natural frequency and the tube's resonant frequency\citep{Cottingham2011}.

Free-reed instruments fall into two families, and the valve configuration follows from how the reed is built. Western free reeds---reed organs, harmonicas, and accordions---have a tongue made separately and mounted on the outside of the plate, and belong to the $(-,+)$ blown-closed class in Fletcher's notation\citep[Chap.~13]{Fletcher1998}. Asian free reeds---\emph{sheng}, \emph{sho}, \emph{khaen}, \emph{bawu}, and \emph{hulusi}\citep{Mores2025} have tongue and frame cut from one piece, belong to the $(+,-)$ blown-open class, and sound only when coupled to a tube. Section~II sets out the two configurations and the onset conditions that separate them. The behavior of one family therefore says little about the other, and the great majority of previous experimental work on free reeds has concerned the Western one.

In addition to fingering, input pressure is another important parameter. Colinot\citep{Colinot2020} measured the vibration of an alto-saxophone reed under fully realistic playing conditions using a single-point optical probe, defined several oscillation regimes including the standard, inverted, and double regimes, and demonstrated that these regimes appear at different blowing-pressure thresholds and can be identified from bifurcation diagrams. The object of that study was a clarinet-like beating reed with a $(-,+)$ configuration, different from a free reed; nevertheless, it showed that the supply pressure itself can change the oscillation regime, a question that has not yet been examined in free-reed instruments.

Cottingham\citep{Cottingham2013} applied mechanical force to accordion reeds using a mechanical exciter and obtained the averaged vibration spectra and response curves of the reeds using a laser vibrometer or a variable-impedance proximity sensor, from which the modes were identified. They further indicated that the first torsional mode and the second transverse mode may play important roles during the initial stage of oscillation. Ukshini\citep{Ukshini}, in contrast, used a full-field quantitative method to measure the three-dimensional vibration of an alto-saxophone reed under airflow excitation, artificial lip force, and complete-instrument conditions. These two studies represent two respective approaches to vibration measurement of free reeds and beating reeds: the former concerns Western free reeds not coupled to a tube, whereas the latter, although measured under realistic playing conditions, concerns a beating reed. Spatially resolved vibration measurements of Asian free reeds coupled to a tube remain absent.

Although the acoustic mechanisms of free-reed instruments have been discussed in several studies, knowledge specifically concerning the hulusi remains extremely limited. The hulusi is a transposing instrument, commonly made in the keys of F, B$\flat$, and C. In the present study, the F-key hulusi covers C4--D5, the B$\flat$-key hulusi covers F4--G5, and the C-key hulusi covers G4--A5.

The instrument carries behavior familiar from nonlinear dynamics\citep{Bader2013}. At normal playing pressure the tube sets the pitch, and opening or closing the finger holes changes it, so the instrument is close to a flute in this respect. At low pressure the pitch moves a long way, and the finger holes change the timbre instead of the pitch: the system has two stable regimes. Normal pressure is how the instrument is played, but the pressure passes through the low range during the initial transient, so the low-pressure regime enters the attack and with it the character of the sound. The attack is led by the reed, which is why the transient is worth describing precisely, and it is in the attack that the \emph{hulusi} departs from other reed instruments and stands with the Chinese \emph{sheng} and the Japanese \emph{sho}\citep{Hikichi2003}.

This paper examines a sound-generating mechanism for which spatially resolved measurements have so far been lacking: the Asian free reed coupled to a tube. One aim is the temporal behavior of the coupled state---how it is established, under what conditions it breaks down, and to what state the reed returns afterwards. The other is the extent to which the reed and the air column dominate the oscillation under simulated playing conditions: whether their frequencies can be separated, how their difference varies with fingering and supply pressure, and whether the relationship holds across instruments. The approach is experimental. Measurements were performed on six hulusi, covering the keys of F, B$\flat$, and C with two instruments in each key, to establish whether the observed relationships belong to individual instruments or to this configuration; two air-supply conditions separate the effects of pressure and air-column loading. Reed vibration was measured by point-by-point scanning, so that the spatial vibration pattern of each frequency component could be reconstructed rather than only its frequency.

The paper addresses the following questions:

\begin{itemize}
    \item How is the coupled state established, and under what conditions does it break down? To what state does the system return after the breakdown?

    \item Is the reed-dominant component that coexists with the air-column frequency during steady-state oscillation the vibration of the reed itself? How does it vary with fingering and reed-side pressure?

    \item Does nonlinear amplitude modulation occur under the low air-supply condition?

    \item What are the spatial vibration patterns of the two oscillation regimes, and how do they change with air-column loading?
\end{itemize}

The paper is organized as follows. Section~II presents the framework of reed--tube coupling and defines the frequency quantities used in this study; Sec.~III describes the instruments, measurement setup, and signal processing; Sec.~IV reports the results: regime transition and recovery of the reed-dominant state (A), frequency relationships and pressure dependence (B), amplitude modulation at low pressure (C), and operating vibration patterns (D); Sec.~V presents the discussion, and Sec.~VI gives the conclusions.

\section{\label{sec:2} Reed--Tube Coupling Framework}

The sound-generating element of the \emph{hulusi} is a free reed: the
reed tongue is slightly smaller than the opening in its frame and swings
back and forth through that opening, in the manner of a swinging door\citep{Cottingham2011}.

The condition for self-sustained oscillation can be written as a balance of acoustic admittances\citep[Eq.~(15.24)]{Fletcher1998}:

\begin{equation}
Y_r(\omega) + Y_s(\omega) + Y_p(\omega) = 0.
\label{eq:admittance_balance}
\end{equation}

where $Y_r$ is the mechanical admittance of the reed, $Y_s$ is the acoustic admittance of the reed opening, and $Y_p$ is the input admittance of the downstream air column. The condition $\mathrm{Re}(Y_r) < 0$ corresponds to energy input, whereas the balance of the imaginary parts sets the oscillation frequency. This equation has more than one solution, and which branch carries the oscillation depends on the magnitude of the air-column impedance at the candidate frequencies.

Valve configurations are classified by Fletcher and Rossing using the binary notation $(\sigma_1,\sigma_2)$: $\sigma_1 = +1$ indicates that the upstream pressure tends to open the valve, whereas $-1$ indicates that it tends to close the valve; $\sigma_2$ is defined analogously for the downstream pressure. Clarinet and double reeds belong to the $(-,+)$ class, corresponding to what Helmholtz termed inward-striking reeds, also referred to as blown-closed configurations; brass-player lip valves belong to the $(+,-)$ outward-striking class or the $(+,+)$ class\citep[Chap.~13]{Fletcher1998}\citep{Helmholtz1954}.

Free-reed instruments fall into two families, with the configuration determined by the construction of the reed\citep{Cottingham2011}. Western free reeds (reed organs, harmonicas, and accordions) have reed tongues fabricated separately and mounted on the outside of the reed plate. At rest, the opening faces the high-pressure side, and the pressure difference drives the tongue toward the frame, they therefore belong to the $(-,+)$ blown-closed configuration. Their sounding frequency is lower than the natural vibration frequency of the reed tongue, and they can usually sound without a tube resonator. Asian free reeds (\emph{sheng, sho, khaen, bawu,} and \emph{hulusi}) have the reed tongue and frame cut from the same piece of material and are approximately symmetric with respect to the airflow direction. The tongue does not lie flush with the plate: a permanent set holds its free end clear of the opening at rest, by $h_0 = 490~\mu$m on the reed measured here (Table~I, Sec.~III~D). The flow therefore meets an opening that is already there, and the pressure difference carries the tongue further from the plate, these reeds belong to the $(+,-)$ blown-open configuration. Such reeds must be coupled to a tube to sustain stable oscillation, and the sounding frequency of the reed--tube combination is higher than both the natural frequency of the reed and the tube resonance frequency\citep{Cottingham2009}. The \emph{hulusi} measured in the present study belongs to the latter family.

The linearized onset conditions are given separately for the two configurations\citep[Eq.~(13.8)]{Fletcher1998}: the $(-,+)$ configuration requires $\omega < \omega_r$ and $X_1 + X_2 > 0$, whereas the $(+,-)$ configuration requires $\omega > \omega_r$ and $X_1 + X_2 < 0$. The blown-open configuration therefore predicts that the coupled sounding frequency lies above the reed frequency, a relationship verified in Sec.~IV.A in the form $\Delta > 0$.

Cottingham expressed the frequency-selection condition in terms of phase matching: at the oscillation frequency, the phase lag of the reed response and the phase of the air-column impedance must cancel, i.e.,
$\tan\theta = -\tan\varphi$\citep{Cottingham2009}. When the air-column resonance is far from the reed's own frequency, this condition can be met near two frequencies, and the system may then select between a reed-dominant frequency and an air-column-dominant frequency.

Three frequencies are used in what follows. The quantity $f_r$ is the mechanical resonance frequency of the reed without air-column loading, $f_p$ is the frequency of the first input-impedance peak of the air column for a given fingering, and $f_c$ is the actual self-sustained oscillation frequency. Two quantities are measured: the spectral peak locked to the air column within the steady-state window, denoted $f_\mathrm{tube}$, and the spectral peak within the onset window that is not in an integer ratio with it, denoted $f_\mathrm{reed}$. Their difference and ratio,

\begin{equation}
\begin{aligned}
\Delta &= f_{\mathrm{tube}} - f_{\mathrm{reed}},\\
R &= \frac{f_{\mathrm{tube}}}{f_{\mathrm{reed}}}.
\end{aligned}
\label{eq:detuning_ratio}
\end{equation}

are computed for each scan and then averaged. Expressed in cents, $R$ becomes $1200\log_2 R$, the musical interval between the reed and the air column. In this study, $\Delta$ is used as the organizing parameter, with $\Delta \to 0$ corresponding to the frequency-capture state in which the two components become indistinguishable.

\section{\label{sec:3} Experimental Methods}

\subsection{Instruments and Reeds}

Six \emph{hulusi} were measured in this study, covering three keys with two instruments in each key (Table~I). The instruments were obtained from instrument makers in Yunnan and were not modified. For each \emph{hulusi}, the reed of the main tube (melody pipe) was measured, with the two side pipes removed. The reed tongue and frame are cut from the same brass sheet, and the tongue carries a permanent set that holds its free end clear of the plate [Fig.~1(b), Sec.~III~D]. Reed dimensions differ among keys (Table~I) and were measured under the microscope [Fig.~1(c)].

Fingerings are named using Chinese solmization (usol, ula, usi, do, re, mi, fa, sol, la), with nine fingerings for each instrument covering approximately one octave. Throughout the paper, instruments are identified using Roman numerals in parentheses [F(i), F(ii), B$\flat$(i), etc.] to avoid confusion with scientific pitch notation (C4, A5).

\begin{table}[!t]
\centering

\caption{The six \emph{hulusi}. Identifiers use roman numerals in parentheses so
that they cannot be read as scientific pitch names. The pitch range is
that covered by the nine fingerings of the melodic pipe. $N$ is the
number of repeat scans at each of the 29 measurement points, for each
fingering and each supply condition. Reed dimensions are of the
melodic-pipe tongue: $L$ is the tongue length, and $W_\mathrm{root}$ and
$W_\mathrm{tip}$ the widths at the clamped and free ends, all three
measured under the microscope [Fig.~1(c)] and used to lay out the scan
grid. The tip offset $h_0$ is the height of the free end above the
surrounding plate, measured by confocal microscopy (Fig.~4) on the one
instrument that was scanned; taking the plate as the reference makes
$h_0$ independent of how the specimen sat on the stage. Tongue and frame
are cut from a single brass sheet, and the tongue carries a permanent set
that lifts its free end clear of the plate. The tongue dimensions carry
an uncertainty of $\pm0.02$~mm, $h_0$ one of $\pm5~\mu$m.}
\label{tab:instruments}

\vspace{0.3em}

\begingroup
\setlength{\tabcolsep}{2.0pt}
\renewcommand{\arraystretch}{1.10}
\scriptsize

\resizebox{\columnwidth}{!}{%
\begin{tabular}{@{}llcccccccc@{}}
\hline\hline

\shortstack{Instrument}
& \shortstack{Key}
& \shortstack{Pitch\\range}
& \shortstack{Finger-\\ings}
& $N$
& \shortstack{$L$\\(mm)}
& \shortstack{$W_\mathrm{root}$\\(mm)}
& \shortstack{$W_\mathrm{tip}$\\(mm)}
& \shortstack{$h_0$\\($\mu$m)}
& \shortstack{$f_\mathrm{reed}$\\(Hz)}
\\
\hline

F(i)         & F        & C4--D5 & 9 & 10 & 18.37 & 2.68 & 0.70 & 490 & 263 \\
F(ii)        & F        & C4--D5 & 9 &  3 & 19.04 & 2.57 & 0.71 & -- & 259 \\
B$\flat$(i)  & B$\flat$ & F4--G5 & 9 &  3 & 15.19 & 1.91 & 0.55 & -- & 342 \\
B$\flat$(ii) & B$\flat$ & F4--G5 & 9 &  3 & 14.63 & 2.24 & 0.56 & -- & 345 \\
C(i)         & C        & G4--A5 & 9 &  3 & 14.49 & 2.07 & 0.56 & -- & 383 \\
C(ii)        & C        & G4--A5 & 9 &  3 & 14.09 & 2.34 & 0.62 & -- & 384 \\

\hline\hline
\end{tabular}%
}

\endgroup
\end{table}

\begin{figure*}[!t]
    \centering
    \includegraphics[width=\textwidth]{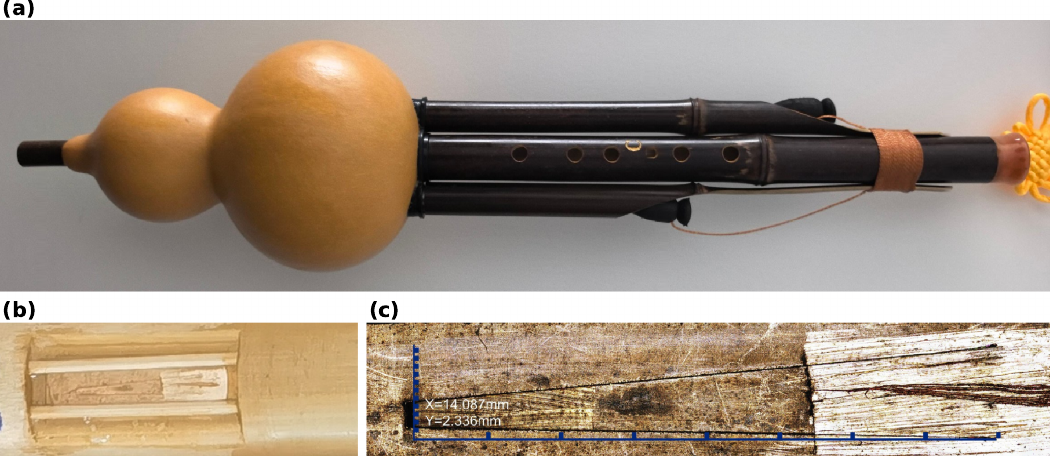}
    \caption{The instrument. (a) A complete \emph{hulusi}: air enters at
    the mouthpiece, the gourd is the wind chamber, and the three pipes
    set into it are the melodic pipe, which carries the finger holes,
    and two drones. (b) The reed of C(ii) in the window cut in the wall
    of its socket. (c) The tongue of the same reed under the microscope,
    with the measurement overlay of the instrument: the tongue is
    14.087~mm long and 2.336~mm wide at the clamped end, which is at the
    left. These are the $L$ and $W_\mathrm{root}$ of Table~I.}
    \label{fig:FIG1}
\end{figure*}

\begin{figure}[!t]
    \centering
    \includegraphics[width=\columnwidth,keepaspectratio]{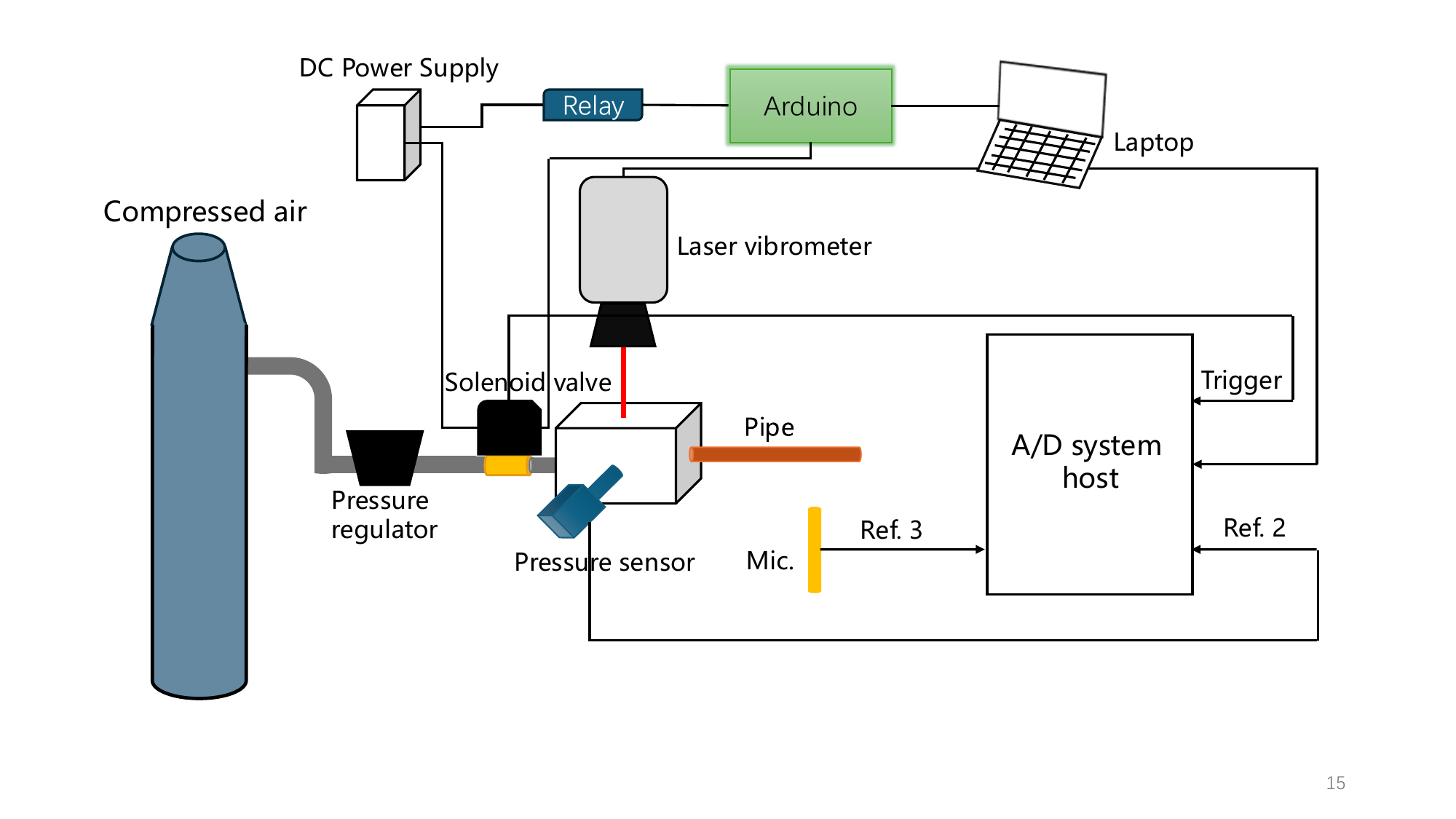}
    \caption{Schematic of the simulated playing rig. Compressed air
    passes from the reservoir through a regulator to a solenoid valve,
    driven from a DC supply through a relay and switched by an Arduino on
    a fixed 1000~ms cycle, and enters one end of a sealed Perspex chamber
    holding the reed. The pressure transducer (Ref2) is on the chamber
    wall, level with the inlet and on the far side of the reed. The laser
    enters vertically from above. The melodic pipe is inserted at the end
    opposite the inlet and passes out of the chamber, leaving the pipe
    outside and the reed inside. The reference microphone (Ref3) is at
    30~cm. All three channels share one trigger.}
    \label{fig:FIG2}
\end{figure}

\subsection{Simulated Playing Setup}

Measurements were conducted under simulated playing conditions (Fig.~2). Compressed air was supplied from an air reservoir through a pressure regulator and switched on and off by a solenoid valve driven by a DC power supply through a relay. The relay was controlled by an Arduino, and a fixed 1000~ms timing sequence was sent by the host computer, so that the air-supply timing was fully repeatable for every scan.

The airflow entered one end of a sealed acrylic chamber from the solenoid valve, with the reed located inside the chamber. The pressure sensor (Ref2) was mounted on the chamber wall at the same horizontal level as the air inlet and on the opposite side of the reed, and therefore measured the pressure at the reed rather than the supply pressure. The laser was incident vertically from above the chamber onto the reed tongue. The main tube of the hulusi was inserted into the chamber from the end opposite the air inlet and extended outside the chamber, so that the tube body remained outside while the reed was inside, allowing the laser to illuminate the reed directly while maintaining a sealed airflow path.

Three signal channels were acquired synchronously: Ref1 was the velocity output of the laser vibrometer, Ref2 was a pressure sensor located downstream of the reed, and Ref3 was a reference microphone positioned 30~cm from the instrument. All three channels shared the same trigger, so the phase relationships within the same event were available.

\subsection{Point-by-Point Laser Vibrometry and Coordinate Reconstruction}

Vibration was measured point by point using a single-point laser Doppler vibrometer. For each fingering and condition, 29 measurement points were placed on the reed tongue (Fig.~3): three scan lines were arranged along the longitudinal axis of the reed, with ten stations on each line and one station omitted at the tip because of insufficient width. Each point was acquired repeatedly 3--10 times, with 10 repetitions for F(i). The sampling rate was 51.2~kHz, and each recording was 1000~ms long.

\begin{figure}
    \centering
    \includegraphics[height=0.7\columnwidth]{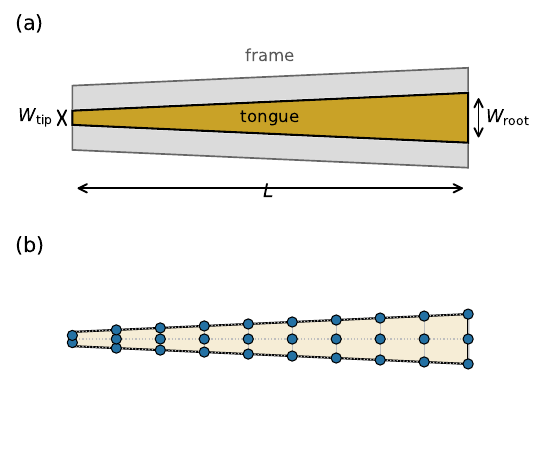}
    \caption{Reed geometry and scan grid, F(i). (a) Tongue and frame to
    scale, with the quantities of Table~I marked: tongue length $L$ and
    the widths $W_\mathrm{root}$ and $W_\mathrm{tip}$ at the clamped and
    free ends. (b) The 29 measurement points: three lines along the
    tongue, ten stations on each, one station omitted at the tip where
    the tongue is too narrow to carry three.}
    \label{fig:FIG3}
\end{figure}

The coordinates obtained from the vibrometer's built-in three-dimensional alignment function placed several measurement points outside the actual scan lines, which deformed the interpolated reed contour. The scan topology was known---three lines, ten stations per line, one station omitted at the tip, and 29 points in total---so we built an ideal trapezoidal grid from the tongue length and the root and tip widths of Table~I, discarded the alignment coordinates, and kept only the vibration data and their original order. These three dimensions were measured under the microscope [Fig.~1(c)].

The ordering of the measurement points was not fixed between measurements: scanning sometimes began from the fixed end and sometimes from the middle, so the point number itself was not a position label. The reed orientation was taken from the data---the amplitude falls monotonically from the free end toward the fixed end, and the end carrying the largest velocity amplitude is the tip. Mapped in this way, the 29 points give an operating vibration pattern that is continuous across the reconstructed grid.

The positional uncertainty of the grid follows from the $\pm0.02$~mm accuracy of the reed dimensions and from the point-to-point repeatability of the vibrometer.

\subsection{Three-Dimensional Reed Morphology}

The static geometry of the reed was measured using a confocal microscope (Olympus OLS5000). A height map of $17.28 \times 3.46$~mm was acquired for the melodic-pipe reed of F(i), with resolutions of 1.4256~$\mu$m, 1.4360~$\mu$m, and 0.0104~$\mu$m in the lateral, longitudinal, and height directions, respectively.

The scanned field includes both the reed tongue and the surrounding flat reed plate. In each row, the reed tongue appears as a raised band with a flat top and nearly vertical sides. Its two boundaries were therefore determined row by row, and three profile lines were placed within the band following its taper. All heights were measured relative to the plate; the inclination of the reed on the microscope stage affects both surfaces equally and is eliminated by subtraction.

The elevation decreases linearly with distance from the free end, and the three profiles coincide [Fig.~4(b)]. The linear fit gives $R^2 = 0.9994$, with a root-mean-square residual of 3.3~$\mu$m and an elevation range of 454~$\mu$m; the deviations of the individual profiles from their respective straight lines remain within $\pm 12$~$\mu$m [Fig.~4(c)]. The free-end elevation is 490~$\mu$m, the elevation reaches zero at a distance of 9.78~mm from the free end, and the set angle is $2.82^\circ$. The reed tongue is straight and is raised as a whole by a bend near the root.

Beyond the bend, the reed tongue is flush with the plate and is not distinguishable in the height map, but the gap between them can still be traced by the lateral gradient to approximately 15~mm from the free end, where the tongue is approximately 1.8~mm wide. The clamping line lies outside the scanned field; $L$ in Table~I was measured under the microscope [Fig.~1(c)]. The reed tongue is already positioned above the plate at rest, and the free-end elevation $h_0$ is listed in Table~I.

\begin{figure}[!t]
    \centering
    \includegraphics[
        width=\columnwidth,
        keepaspectratio
    ]{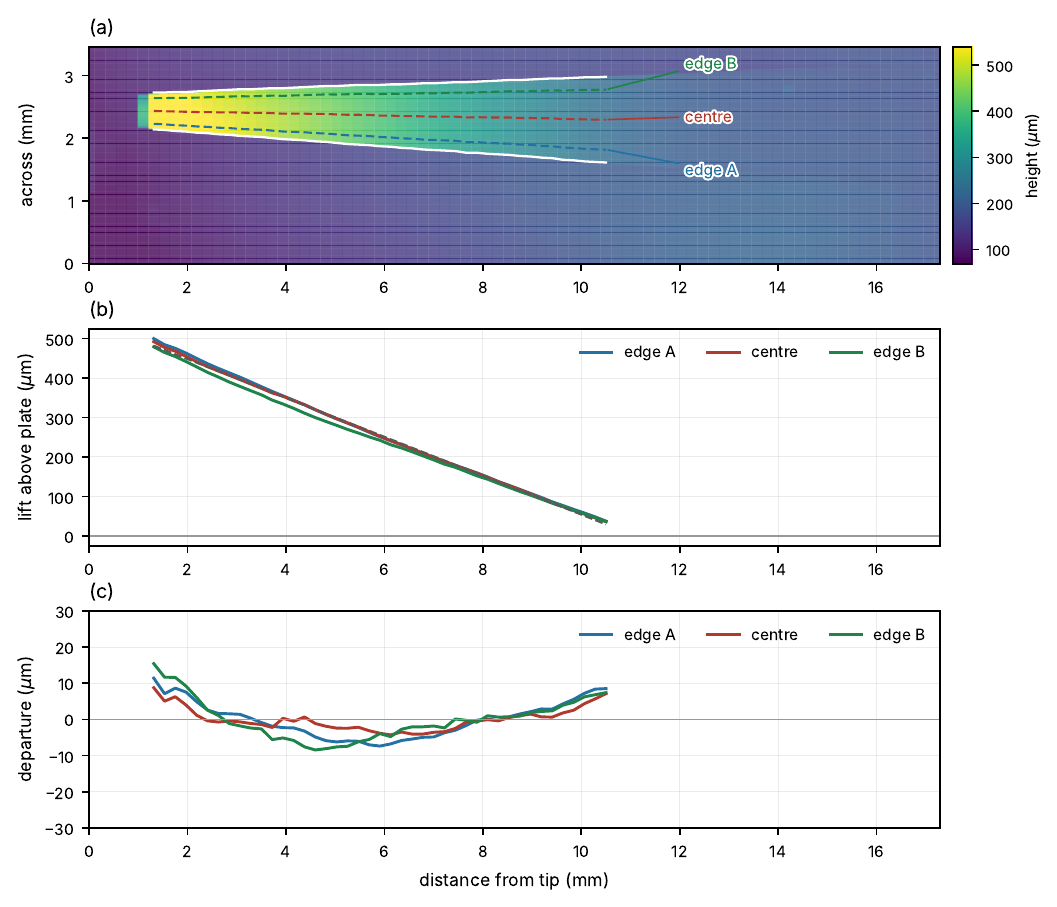}
    \caption{Static geometry of the melodic-pipe reed tongue of F(i),
measured by confocal microscopy. The free end is at the left; the across
axis of (a) is stretched by a factor of 1.28. White lines: tongue
outline. Coloured dashed lines: the three profiles, at 15, 50 and 85 per
cent of the local tongue width. (a) Height over the scanned field,
(b) lift of the tongue above the surrounding plate, with the best
straight line through the three profiles (grey dashed), (c) departure of
each profile from its own best straight line.}
\label{fig:FIG4}
\end{figure}

\subsection{Signal Processing}

\textbf{Segmentation.}
Each recording was divided into onset, coupled steady-state, silent, and reed-dominant segments according to the RMS envelope and its second derivative; the segment boundaries were recorded together with the pressure curve. Frequency extraction was performed within the corresponding windows: $f_\mathrm{tube}$ was obtained from the steady-state window, and $f_\mathrm{reed}$ from the onset window before the establishment of coupling.

\textbf{Frequency extraction.}
A Fourier transform was performed for each scan within each window, and spectral peaks were extracted. The quantity $f_\mathrm{tube}$ was taken as the strongest peak in the steady-state window together with its harmonic series; $f_\mathrm{reed}$ was taken as the strongest peak in the onset window that was not in an integer ratio with $f_\mathrm{tube}$. Two types of scans were excluded from the analysis: those in which the air-column resonance and reed frequency were too close to be resolved within the window, denoted as frequency capture, and those in which no reed component was detected. The search band, analysis-window length, and peak criteria for each instrument were adjusted individually according to its reed frequency and signal-to-noise ratio; the parameter values are listed in the Supplementary Material.

\textbf{Operating vibration patterns.}
At each measurement point, the complex amplitude near the target frequency was extracted from the velocity spectrum. Collecting these over the 29 points gives the vibration pattern of the reed at that frequency, the quantity modal testing calls the operating deflection shape\citep{Richardson1997,Bae2011}: a complex vector $\boldsymbol{\psi}$ with one entry per point. Multiplying $\boldsymbol{\psi}$ by a complex number scales the whole pattern and rotates its phase without changing the motion it describes, so the overall factor carries no information. Each pattern was therefore rotated until the amplitude-weighted complex mean of the tip column was positive and real, which fixes a common starting phase and lets the sign distributions be read across fingerings and instruments.

\textbf{Shape comparison.}
Two patterns were compared by the modal assurance criterion\citep{Allemang1982,Allemang2003}
\begin{equation}
\mathrm{MAC}(\boldsymbol{\psi}_a,\boldsymbol{\psi}_b)
=
\frac{\left|\boldsymbol{\psi}_a^{\mathrm{H}}\boldsymbol{\psi}_b\right|^{2}}
{\left(\boldsymbol{\psi}_a^{\mathrm{H}}\boldsymbol{\psi}_a\right)
\left(\boldsymbol{\psi}_b^{\mathrm{H}}\boldsymbol{\psi}_b\right)},
\label{eq:mac}
\end{equation}
where $^{\mathrm{H}}$ is the conjugate transpose. The normalization cancels any complex factor applied to either pattern\citep{Vacher2010}, so Eq.~(\ref{eq:mac}) measures how the motion is distributed over the reed and not how large it is. It returns 1 for two patterns that differ only by such a factor and 0 for orthogonal patterns. Two unrelated patterns do not give 0 but $1/N$, which is $0.034$ for the $N = 29$ points used here, this is the level against which a measured MAC is read.

Torsion about the long axis was measured separately. The three scan lines are placed symmetrically, so every point $i$ on an off-axis line has a mirror partner $i'$ on the other. The sum $\psi_i + \psi_{i'}$ moves the two sides together and is bending; the difference $\psi_i - \psi_{i'}$ moves them in opposition and is torsion. Their relative weight, defined here as the antisymmetric energy fraction, is
\begin{equation}
\frac{E_a}{E}
=
\frac{\sum_i \left|\psi_i - \psi_{i'}\right|^{2}}
{\sum_i \left(\left|\psi_i + \psi_{i'}\right|^{2}
+ \left|\psi_i - \psi_{i'}\right|^{2}\right)},
\label{eq:anti_fraction}
\end{equation}
which is 0 for pure bending and 1 for pure torsion. Equations~(\ref{eq:mac}) and (\ref{eq:anti_fraction}) answer two different questions about a measured pattern: whether it repeats, and whether it is the bending pattern a cantilever is expected to take.

\textbf{Phase reference for the decoupling segment.}
When the supply pressure is applied, the reed passes through the four segments in turn. During onset it vibrates near its own frequency and the amplitude grows; this window gives $f_\mathrm{reed}$. In the coupled steady state the oscillation is locked to the air column and the amplitude is largest; this window gives $f_\mathrm{tube}$. Locking is then lost and the amplitude falls to about 47~dB below the steady-state level. Sound reappears at the reed frequency, and this reed-dominant segment carries the vibration patterns of Sec.~IV~D.

That collapse cuts the reappearing segment off from everything before it. The second onset grows out of noise, and the phase of the noise that seeds it differs from one triggering event to the next. The velocity phase recorded at a point therefore depends on which seed started that particular scan, and phases measured at different points cannot be placed on a common clock.

A quantity recorded at the same instant as the velocity removes the seed. The microphone signal $m_i$ was acquired at every point together with the velocity $v_i$, and the phase of the transfer function
\begin{equation}
H_i(f)
=
\frac{S_{v_i m_i}(f)}
{S_{m_i m_i}(f)}
\label{eq:transfer}
\end{equation}
was used in place of the velocity phase. Equation~(\ref{eq:transfer}) is the $H_1$ estimator\citep{Bendat2010}, taken here with the microphone as the reference channel\citep{Ukshini}. The seed enters $v_i$ and $m_i$ alike and cancels in the ratio, leaving the phase of the reed relative to the sound it radiates, which is the same quantity at every point. Points whose ordinary coherence\citep{Bendat2010} with the reference fell below 0.5 were excluded from the vibration pattern.

\section{Results}
\label{results}

\subsection{\label{subsec:4:1} Regime Transition and Recovery of the Reed-Dominant State: The B$\flat$(i) Case}

This section follows one recording in time; the statistics over all six instruments follow in Sec.~IV.B. The two spectral components are those defined in Sec.~II: the tube-dominant frequency $f_\mathrm{tube}$, locked to the air-column resonance and carrying a complete harmonic series, and the reed-dominant frequency $f_\mathrm{reed}$, which is not an integer multiple of it. The case is the bbfa fingering of B$\flat$(i), whose air-column resonance is 622.2~Hz under the low-supply condition (Fig.~5). The vibration velocity measured by laser vibrometry and the input pressure together separate the stages of oscillation and show how the coupled state is established and lost. The remaining fingerings of B$\flat$(i) are summarized in Fig.~6.

Throughout this work, ``decoupling'' refers to the loss of frequency locking of the oscillation to the air-column resonance, rather than the disappearance of acoustic coupling---the tube always constitutes a load on the reed. The oscillation segment after the loss of locking is termed the reed-dominant state. This segment remains driven by airflow ($p_\mathrm{reed}$ is 4.5 times the reed oscillation threshold), and its amplitude grows exponentially ($\tau = 29.7$~ms, $R^2 = 0.96$, corresponding to 10.3 cycles).

The spatially averaged velocity waveform [Fig.~5(a)] exhibits a four-stage structure. After the air supply is switched on, the reed begins to oscillate at 96~ms, with a spectral peak at 346.6~Hz and weak harmonics. At 221~ms, the oscillation locks to the air-column resonance, the velocity envelope rises abruptly to a plateau, and a complete harmonic series appears in the spectrum [blue curve in Fig.~5(d)]. After the air supply is switched off, the coupled state collapses at 676~ms, the RMS envelope falls to the measurement noise floor [Fig.~5(b)] and remains there for 100~ms (34.7 reed cycles), after which the oscillation reappears at 776~ms.

The short-time Fourier spectrum [Fig.~5(c)] shows that the air-column fundamental and its entire harmonic series disappear together during the collapse. The energy fraction in the 622-Hz band is 0.8--0.95 during the coupled steady state, falls to nearly zero during the collapse, and stays low thereafter [right axis of Fig.~5(b)].

The reappearing oscillation [green curve in Fig.~5(d)] is dominated by a single peak at 346.5~Hz with weak harmonics, and its spectral shape matches that of the onset segment. This frequency differs by only 0.1~Hz (0~cents) from the reed-dominant frequency in the onset segment of the same measurement, whereas the reed frequency varies by 71--268~cents across fingerings (Sec.~IV.B). The first air-column resonance for this fingering is 622.2~Hz; 346.5~Hz is not its subharmonic ($622.2/2 = 311.1$~Hz, differing by 187~cents), nor does it correspond to a lower tube resonance. The system therefore completes a closed loop in regime space: reed-dominant $\rightarrow$ tube-dominant $\rightarrow$ reed-dominant. The final and initial states belong to the same regime, and the component identified within the onset window is the reed's own vibration frequency.

The synchronously recorded pressure on the reed side [right axis of Fig.~5(a), uncalibrated] shows that the collapse occurs at $p = 1.08$, which is 5.1 times the reed oscillation threshold of 0.21 and 70\% of the peak pressure. During the silent interval the pressure lies between 1.08 and 0.95, more than four times the oscillation threshold. Too small a pressure difference across the reed therefore does not account for the breakdown of the coupled state or for the silence that follows. Locking is established at $p = 1.50$ on the rising branch and breaks down at $p = 1.08$ on the falling branch. The pressure sweep in the present experiment is not quasistatic, and the transition points carry rate-dependent delays; separating true bistability from dynamic hysteresis requires repeated ramps at controlled rates (Sec.~V.D).

Applying the same analysis to the remaining fingerings of B$\flat$(i) (Fig.~6), the four-stage structure repeats for each fingering, with locking established between 227~ms and 251~ms measured from the start of the recording. Three quantities vary monotonically with $\Delta$: the silent interval increases from 36~ms (D5, $\Delta = 237$~Hz) to 84~ms (G5, $\Delta = 419$~Hz), the tube-dominant segment decreases from 499~ms to 329~ms, and the reed-dominant segment increases from 164~ms to 310~ms (Spearman $\rho = \pm 1.00$, exact permutation $p = 0.083$, $n = 4$). The ratio of the RMS of the post-decoupling reed-dominant segment to that of the coupled steady state also increases with $\Delta$ (1.55 to 3.10). Thus $\Delta$ describes not only the steady-state frequency relationship but also the time the system spends in each of the two regimes.


\begin{figure}[!t]
    \centering
    \includegraphics[
        width=\columnwidth,
        keepaspectratio
    ]{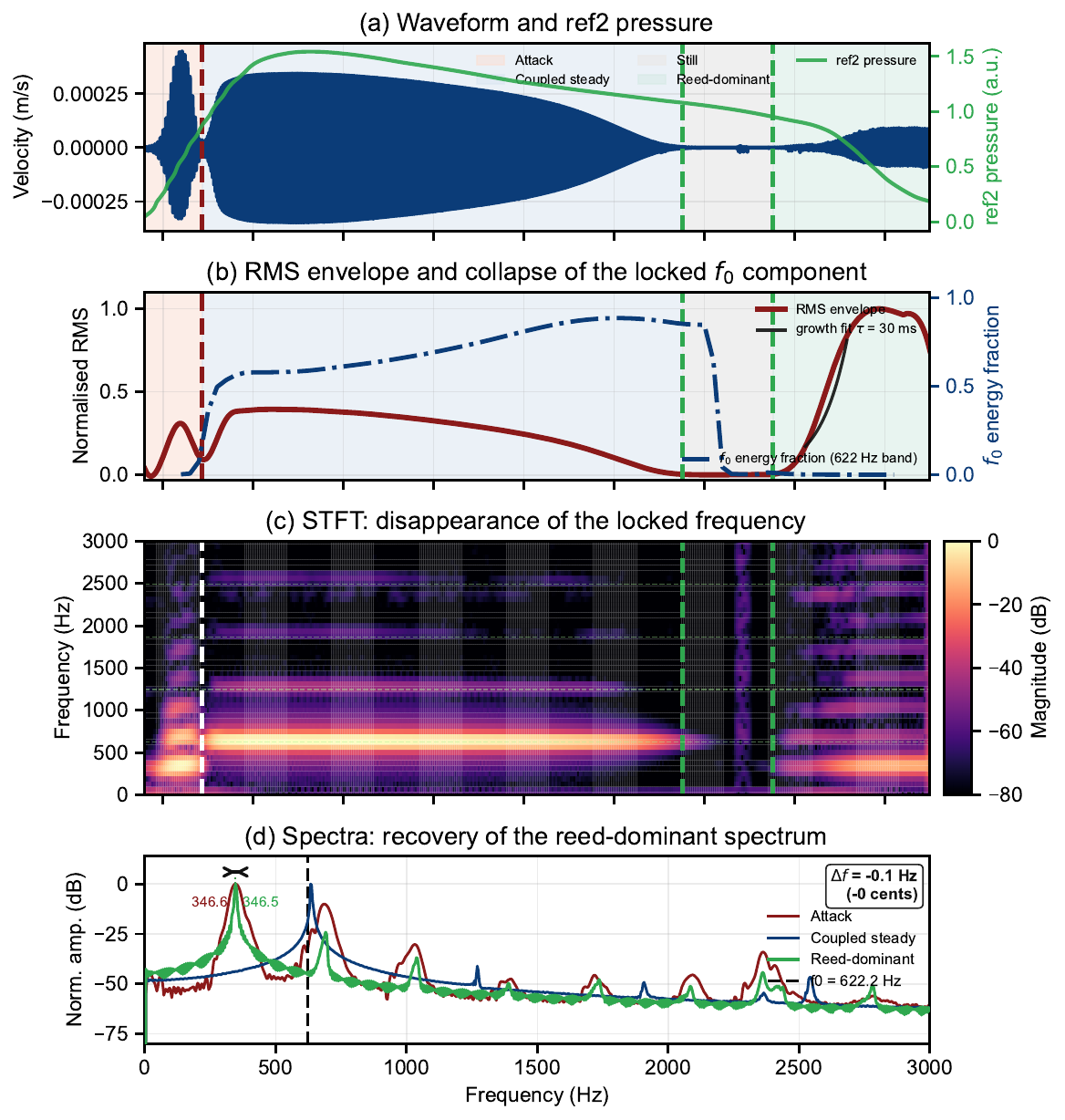}
    \caption{Loss and recovery of lock, B$\flat$(i), E$\flat$5 fingering
    (fa), low supply pressure.
    (a) Velocity averaged over the 29 scan points, with reed-side pressure
    on the right axis (uncalibrated); the reed starts at 96~ms, locks to
    the air column at 221~ms, and the coupled state collapses at 676~ms,
    followed by a 100-ms silent gap and renewed oscillation from 776~ms.
    (b) RMS envelope, with the fraction of energy in the 622-Hz band on
    the right axis.
    (c) Short-time Fourier spectrum; dashed lines mark harmonics of
    $f_{\mathrm{tube}}$.
    (d) Spectra of the attack window, the coupled steady state (blue),
    and the recovered segment (green); the recovered peak (346.5~Hz)
    matches the attack peak (346.6~Hz) to within 1~cent.}
    \label{fig:FIG5}
\end{figure}


\begin{figure*}[!t]
    \centering
    \includegraphics[
        width=\textwidth,
        height=0.30\textheight,
        keepaspectratio
    ]{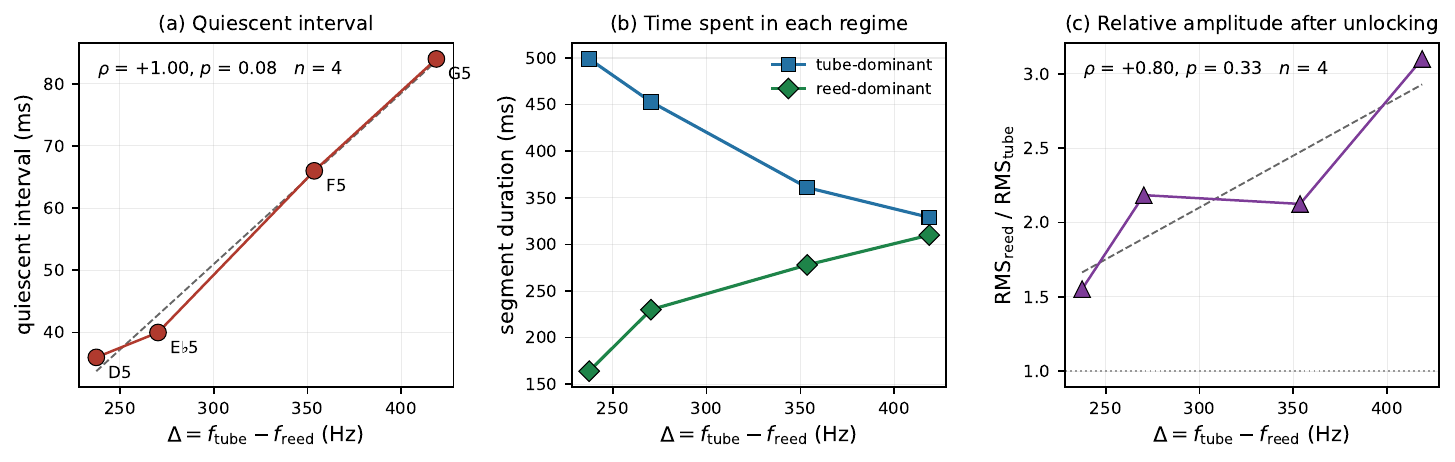}
    \caption{Segment statistics for the four B$\flat$(i) fingerings that
    lose lock, against $\Delta$. (a) Duration of the silent gap;
    (b) durations of the coupled (tube-dominant) segment and of the
    recovered reed-dominant segment; (c) RMS ratio of the recovered to
    the coupled segment. For G5, $\Delta$ is taken from the normal-supply
    measurement.}
    \label{fig:FIG6}
\end{figure*}

\subsection{\label{subsec:4:2}Frequency Relationships, Detuning Staircase, and Pressure Dependence}

The four-stage sequence of Sec.~IV.A is not particular to one instrument. The scan-averaged velocity spectra of steady-state oscillations carry the same two components across all six instruments and both air-supply conditions: $f_\mathrm{tube}$, locked to the air-column resonance and carrying a complete harmonic series, and $f_\mathrm{reed}$, which is not an integer multiple of it. Their difference $\Delta = f_\mathrm{tube} - f_\mathrm{reed}$ and ratio $R = f_\mathrm{tube}/f_\mathrm{reed}$ are recorded for each scan.

Figure~7 presents F(i) as an example. For this instrument, 8--10 scans were acquired for each fingering and condition, with a median inter-scan standard deviation of $p_\mathrm{reed}$ of 0.007~kPa. The tube-dominant frequency follows the fingering scale closely, rising from 297.7~Hz for the D4 fingering to 587.4~Hz for the D5 fingering, with an inter-scan spread below 1~Hz. The reed-dominant frequency instead shifts with fingering: from 251.9~Hz to 268.3~Hz under the low-pressure condition (16.5~Hz, 110~cents) and from 240.2~Hz to 277.4~Hz under the normal-pressure condition (37.4~Hz, 250~cents). Correspondingly, $\Delta$ rises from 45.8~Hz to 324.2~Hz and $R$ from 1.18 to 2.23---the detuning between the reed and the air column widens from about three semitones to slightly more than one octave. The remaining five instruments show the same structure (Fig.~8; individual instrument panels are provided in the Supplementary Material): the span of $f_\mathrm{reed}$ within each instrument ranges from 71~cents to 268~cents, while $\Delta$ runs from 38~Hz to 523~Hz across the entire dataset.

Does the shift in $f_\mathrm{reed}$ come from pressure or from tube loading? The supply pressure is held constant at the air reservoir, but the tube impedance and mean flow resistance vary with fingering: the measured $p_\mathrm{reed}$ of F(i) runs from 1.10~kPa to 1.61~kPa under the low-pressure condition and from 1.28~kPa to 1.84~kPa under the normal-pressure condition.

Changing the fingering alters $p_\mathrm{reed}$ and $f_\mathrm{tube}$ together, their within-unit correlation coefficients running from 0.62 to 0.98. A fixed-effects regression with within-unit demeaning, $f_\mathrm{reed} \sim p_\mathrm{reed} + f_\mathrm{tube}$, separates the two, with standard errors clustered by fingering ($n = 371$ scans, 87 clusters, 11 units; Fig.~8). Pressure keeps a significant partial effect: $\partial f_\mathrm{reed}/\partial p_\mathrm{reed} = +28.0 \pm 10.3$~Hz/kPa ($p = 0.008$), with a partial correlation of $r(p_\mathrm{reed}, f_\mathrm{reed} \mid f_\mathrm{tube}) = +0.33$ ($p < 10^{-4}$). The tube resonance frequency has no detectable effect of its own once pressure is controlled for ($p = 0.52$).

The reed-dominant frequency therefore rises monotonically with the pressure on the reed side, by about 160~cents per kPa (median). This positive dependence follows an outward-striking free reed: a larger pressure difference raises the mean opening and the aerodynamic stiffness term, carrying the self-sustained oscillation frequency upward, opposite to the direction of the inward-striking configuration of a clarinet\citep[Chap.~13]{Fletcher1998}.

\begin{figure*}[!t]
    \centering
    \includegraphics[width=\textwidth]{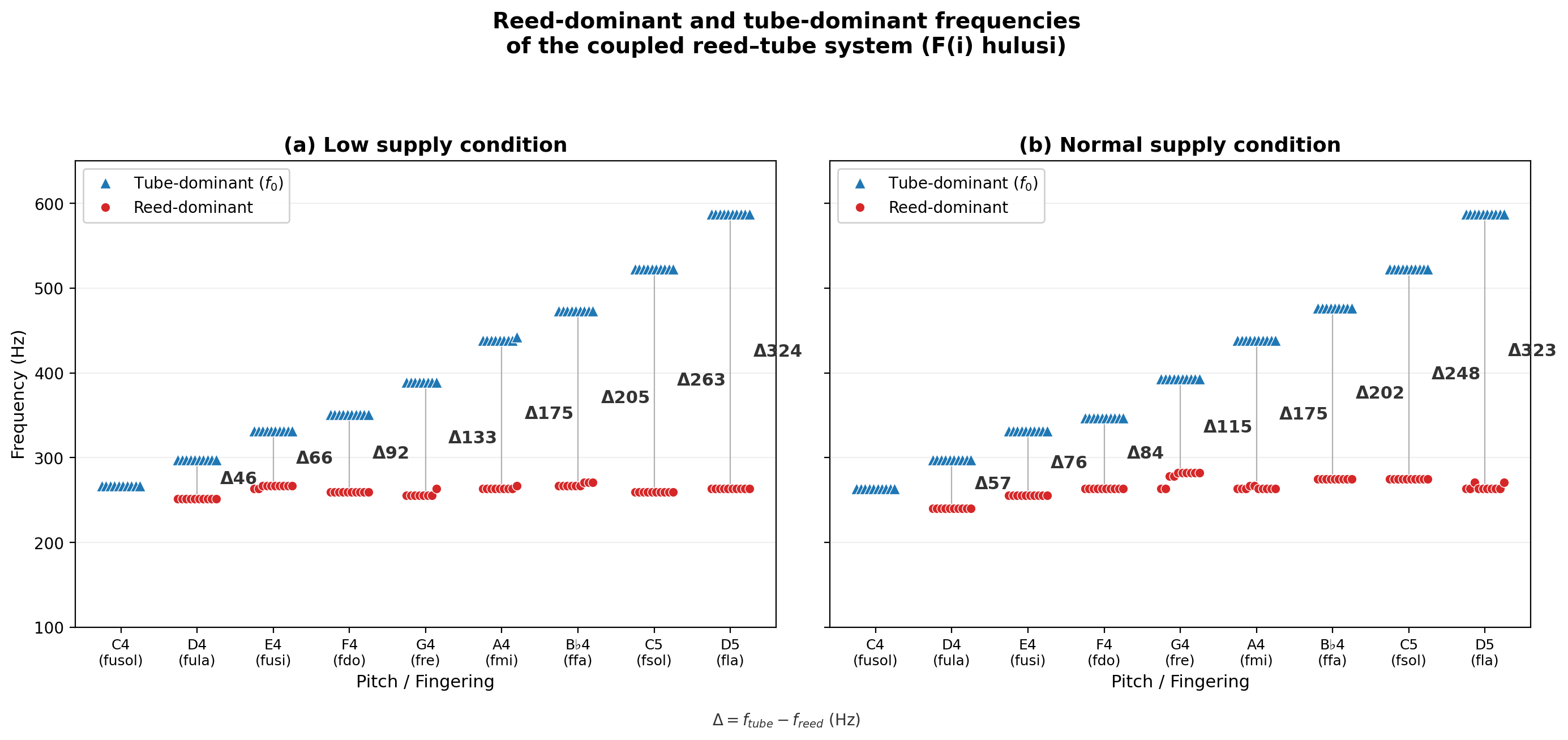}
    \caption{Reed-dominant frequency $f_{\mathrm{reed}}$ (attack window) and
    tube-dominant frequency $f_{\mathrm{tube}}$ of the F(i) hulusi against
    fingering, at (a) low and (b) normal supply pressure. Symbols are the
    mean of 8--10 scans and error bars $\pm 1$ SD across scans. The number
    above each fingering is $\Delta = f_{\mathrm{tube}} - f_{\mathrm{reed}}$
    in Hz. At the C4 fingering the two components cannot be separated
    (frequency capture, $\Delta \to 0$) and only $f_{\mathrm{tube}}$ is
    shown.}
    \label{fig:FIG7}
\end{figure*}

\begin{table*}[!t]
\centering
\caption{Frequency--pressure statistics for individual
instrument--condition units. Each row is one instrument under one supply
condition; the unit within a row is the fingering, and every quantity is
computed across the $n$ fingerings of that row. $r$, Pearson correlation
between the mean supply pressure $P$ and $f_\mathrm{reed}$; CI low and CI
high, its 95\% confidence interval from the Fisher $z$ transform; $p$,
two-sided significance of $r$; $\rho$, Spearman rank correlation. Hz/kPa
and cent/kPa, slope of the least-squares regression of $f_\mathrm{reed}$
on $P$. $r_\mathrm{Pft}$, Pearson correlation between $P$ and
$f_\mathrm{tube}$; $r_\mathrm{part}$ and $p_\mathrm{part}$, partial
correlation of $P$ with $f_\mathrm{reed}$ at fixed $f_\mathrm{tube}$ and
its significance. Span, the range of $f_\mathrm{reed}$ over those
fingerings. Dashes mark C(ii) low, in which no scan passed the quality
criteria.}
\label{tab:frequency_pressure_statistics}

\setlength{\tabcolsep}{3.0pt}
\renewcommand{\arraystretch}{1.15}

\resizebox{\textwidth}{!}{%
\begin{tabular}{ll*{13}{r}}

\hline\hline

\multicolumn{2}{c}{}
& \multicolumn{6}{c}{Frequency--pressure correlation}
& \multicolumn{2}{c}{Regression slope}
& \multicolumn{3}{c}{Partial correlation}
& \multicolumn{2}{c}{Frequency variation}
\\

\cline{3-8}
\cline{9-10}
\cline{11-13}
\cline{14-15}

\shortstack{Instrument}
&
\shortstack{Pressure\\condition}
&
$n$
&
$r$
&
\shortstack{CI\\low}
&
\shortstack{CI\\high}
&
$p$
&
$\rho$
&
\shortstack{Hz/\\kPa}
&
\shortstack{cent/\\kPa}
&
$r_{\mathrm{Pft}}$
&
$r_{\mathrm{part}}$
&
$p_{\mathrm{part}}$
&
\shortstack{Span\\(Hz)}
&
\shortstack{Span\\(cent)}
\\

\hline

F(i) & low
& 8 & 0.606 & -0.172 & 0.918 & 0.1114 & 0.299
& 17.6 & 116.5
& 0.921 & 0.637 & 0.1242
& 16.5 & 110.1
\\

F(i) & normal
& 8 & 0.839 & 0.328 & 0.970 & 0.0092 & 0.611
& 48.6 & 318.2
& 0.835 & 0.771 & 0.0424
& 37.4 & 250.3
\\

F(ii) & low
& 8 & 0.841 & 0.334 & 0.970 & 0.0089 & 0.951
& 17.7 & 119.2
& 0.927 & 0.601 & 0.1539
& 14.0 & 95.0
\\

F(ii) & normal
& 8 & 0.814 & 0.255 & 0.965 & 0.0140 & 0.791
& 10.0 & 67.3
& 0.903 & 0.650 & 0.1137
& 11.4 & 77.0
\\

B$\flat$(i) & low
& 7 & 0.864 & 0.316 & 0.980 & 0.0122 & 0.964
& 31.7 & 160.7
& 0.846 & 0.663 & 0.1516
& 16.5 & 83.5
\\

B$\flat$(i) & normal
& 8 & 0.851 & 0.364 & 0.972 & 0.0074 & 0.619
& 90.8 & 453.3
& 0.653 & 0.892 & 0.0070
& 48.7 & 247.6
\\

B$\flat$(ii) & low
& 8 & 0.518 & -0.293 & 0.896 & 0.1881 & 0.762
& 11.3 & 56.3
& 0.624 & 0.065 & 0.8898
& 14.4 & 71.4
\\

B$\flat$(ii) & normal
& 8 & 0.766 & 0.133 & 0.955 & 0.0267 & 0.857
& 44.8 & 221.8
& 0.750 & 0.260 & 0.5728
& 36.2 & 180.5
\\

C(i) & low
& 8 & 0.564 & -0.233 & 0.908 & 0.1451 & 0.476
& 107.8 & 470.6
& 0.914 & 0.516 & 0.2360
& 62.6 & 268.2
\\

C(i) & normal
& 8 & 0.269 & -0.538 & 0.818 & 0.5202 & 0.000
& 23.9 & 107.1
& 0.921 & 0.785 & 0.0366
& 36.6 & 160.5
\\

C(ii) & low
& 0 & -- & -- & -- & -- & --
& -- & --
& -- & -- & --
& -- & --
\\

C(ii) & normal
& 8 & 0.973 & 0.855 & 0.995
& $4.63\times10^{-5}$ & 1.000
& 35.8 & 159.2
& 0.980 & 0.585 & 0.1680
& 23.8 & 105.7
\\

\hline\hline

\end{tabular}%
}

\end{table*}


\begin{figure}[!t]
    \centering
    \includegraphics[
        width=\columnwidth,
        keepaspectratio
    ]{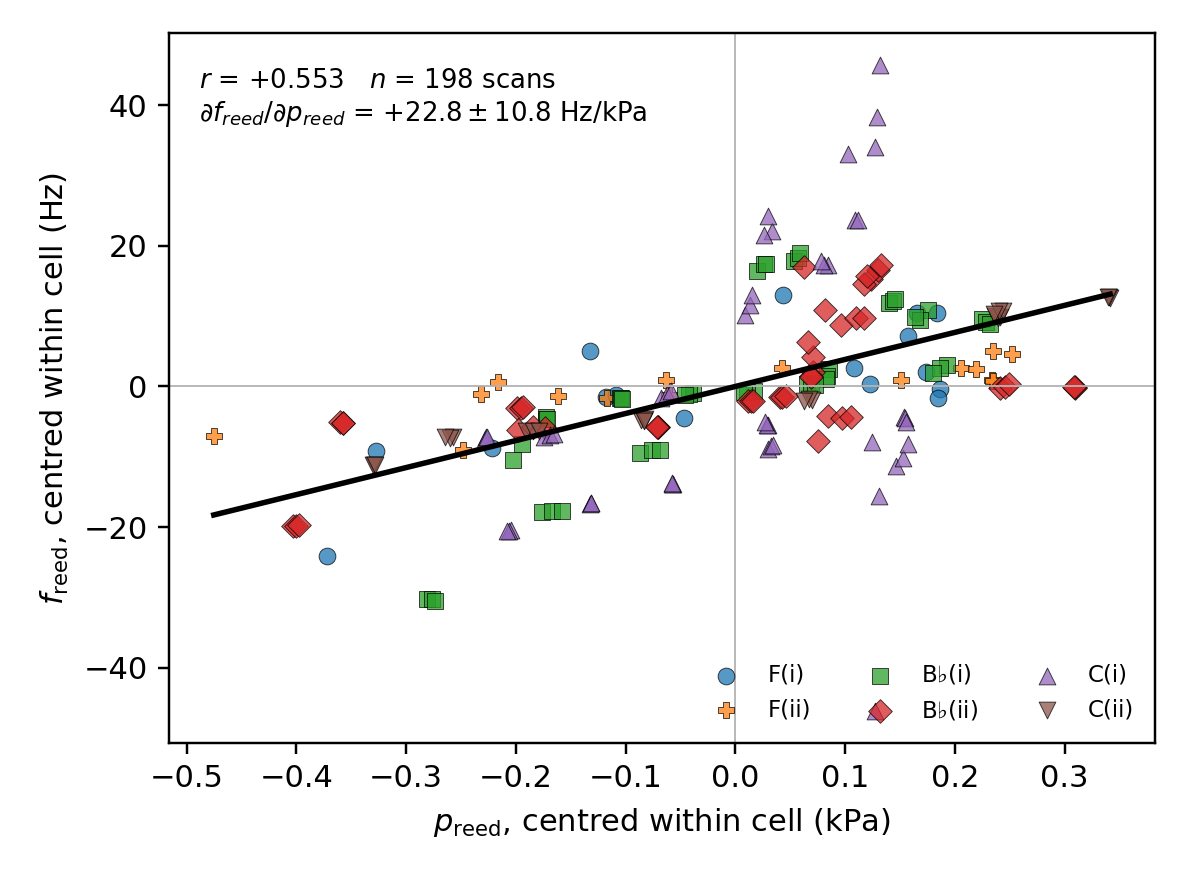}
    \caption{Reed-dominant frequency against reed-side pressure for the
    six instruments, both axes centred on the mean of each instrument
    $\times$ supply-condition cell ($n = 371$ scans, 11 cells). Marker
    shape identifies the instrument. The line is the pooled
    fixed-effects fit controlling for $f_\mathrm{tube}$ (Table~II).}
    \label{fig:FIG8}
\end{figure}

\begin{figure*}[!t]
    \centering
    \includegraphics[
        width=\textwidth,
        height=0.36\textheight,
        keepaspectratio
    ]{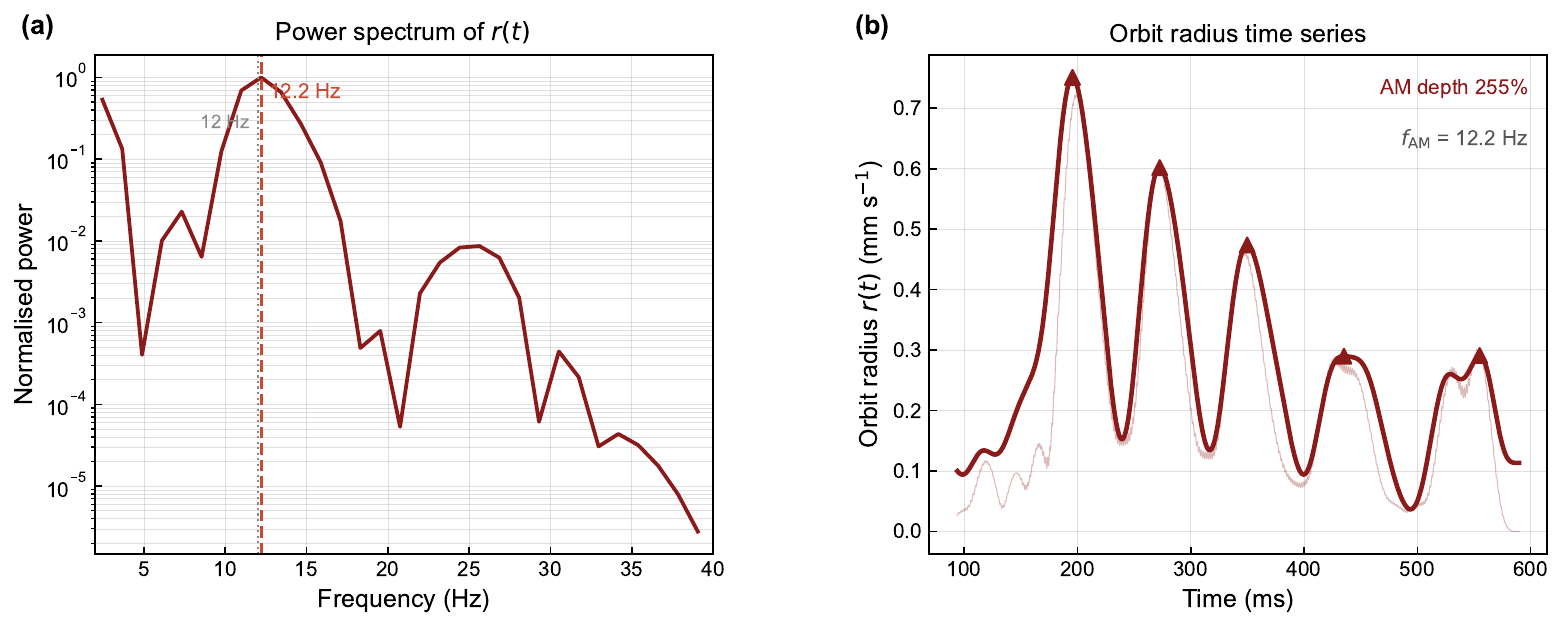}
    \caption{Amplitude modulation of F(ii), C4 fingering, low supply pressure.
    (a) Orbit radius
    $r(t)=\sqrt{v^2+(\dot{v}/\omega_0)^2}$ in the steady window;
    modulation depth 255\%, with inter-peak intervals lengthening
    from 72~ms to 112~ms as the peak amplitude decreases.
    (b) Power spectrum of $r(t)$, peak at 12.2~Hz.
    }
    \label{fig:FIG9}
\end{figure*}

\subsection{\label{subsec:4:3} Amplitude Modulation at Low Pressure}

Under the normal air-supply condition, the steady-state velocity envelopes of all measured fingerings remain flat. In contrast, the lowest fingering C4 (f1usol) of F(ii) behaves differently under the low-supply condition: the dominant oscillation frequency stays locked to the air-column resonance, while the amplitude envelope carries a slow quasiperiodic modulation that lasts through the segment, as shown in Fig.~9.

The oscillation was followed through the instantaneous orbit radius
\begin{equation}
r(t) = \sqrt{v^{2} + \left(\dot{v}/\omega_0\right)^{2}},
\label{eq:orbit_radius}
\end{equation}
where $\omega_0 = 2\pi f_0$ brings $\dot{v}$ to the units of $v$. Equation~(\ref{eq:orbit_radius}) is the radius of the trajectory in the $(v,\dot{v}/\omega_0)$ plane, and it follows the amplitude of the oscillation from one cycle to the next. Within the steady-state window the envelope of $r$ passes through about five maxima, and its power spectrum has one narrow peak at 12.2~Hz [Figs.~9(a) and 9(b)]. The depth of the modulation, taken as $(r_\mathrm{max} - r_\mathrm{min})/r_\mathrm{min}$, is 255\%: the radius at a maximum is about $3.5$ times the radius at the neighbouring minimum.

In phase space the individual $f_0$ cycles trace a family of nested closed loops about a common centre. The outer loops are three to four times the radius of the inner ones, which is the same ratio as the depth, and the loops keep their shape as the radius changes; successive cycles overlap closely. The velocity power spectrum has a single peak at $f_0$. The modulation therefore acts on the radius of one limit cycle. Switching between regimes would leave a second peak, and two superposed components would leave a pair of sidebands; neither appears.

For the D5 fingering (f1la) of the same instrument under the same condition, the envelope stays flat through the steady-state segment, with a depth below 5\%, so the behaviour depends on the fingering. The pressure envelope recorded at the same time stays nearly constant while the modulation runs, which places the source inside the system rather than in the supply.

This coexists with the two-regime structure of Secs.~IV.A and IV.B: the oscillation stays locked to the air-column frequency throughout, and only the amplitude changes. Its origin---weakened transverse stability of the limit cycle near the threshold of a supercritical Hopf bifurcation, together with the relation between the pressure and velocity envelopes---rests on a separate set of measurements and will be reported elsewhere.

\subsection{\label{subsec:4:4} Operating Vibration Patterns}

The preceding three sections discussed the frequency relationship between the reed and the tube; this section describes the spatial motion of the reed vibration. We select the reed of B$\flat$(i), discussed previously in Sec.~IV.A, as a representative case because it exhibits a state in which oscillation reappears after decoupling, allowing comparison with the initial state at onset.

Each reed was sampled along three longitudinal scan lines, with ten stations per line and one station omitted at the tip, giving a total of 29 points (Fig.~3). At each point, the complex amplitude was extracted from the velocity spectrum near the target frequency to construct the operating vibration pattern at that frequency. Because the global phase of a complex vibration pattern is arbitrary, all vibration patterns were rotated to a positive real reference using the amplitude-weighted complex mean of the tip column, so that the sign distributions in the figures can be compared across panels. The vibration patterns at different frequencies are the spatial distributions of the same self-excited limit cycle at the fundamental frequency and its harmonics, rather than structural modes: the reported frequencies are in strict integer ratios (336/671/1003~Hz for B$\flat$(i) under low pressure), whereas the bending-mode frequency ratios of a cantilever reed are far from 1:2:3. We denote the harmonic orders as H1, H2, and H3. The harmonic order is determined from the measured frequency ratio rather than from the extraction label, and spectral peaks whose ratios deviate from an integer by more than 8\% are excluded.

Shapes are compared by the modal assurance criterion of Eq.~(\ref{eq:mac}). Every value quoted in this section is taken between two patterns measured on the same instrument, in the same regime, under the same supply condition, and at the same harmonic; the two differ only in fingering, and we call the result the cross-fingering MAC. The A4 fingering is the reference throughout, so each value is one fingering against A4: $\boldsymbol{\psi}_a$ in Eq.~(\ref{eq:mac}) is the pattern of the fingering in question and $\boldsymbol{\psi}_b$ that of A4. A value near 1 means the reed moves in the same pattern for both fingerings; the level expected of two unrelated patterns is $1/29 = 0.034$.

\textbf{Reed-dominant state.} Figure~10 shows the fundamental vibration patterns of B$\flat$(i) for four fingerings in the two regimes under the two air-supply conditions, with the detuning increasing from 100 to 237~Hz. The two reed-dominant groups exhibit the same pattern: the amplitude decreases monotonically from the reed tip to the root, the sign is uniform over the entire reed, and no nodal line is present. For these four fingerings against A4, the cross-fingering MAC ranges from 0.78 to 0.92 under low pressure and from 0.67 to 0.85 under normal pressure.

Extending the analysis to all six instruments, the median cross-fingering MAC of the fundamental in the reed-dominant state ranges from 0.58 to 0.998 (Table~III). F(i) is the highest: there even the lowest single comparison reaches 0.99. The reed-dominant vibration pattern therefore changes little with fingering or air-supply condition.

\begin{table*}[!t]
\centering
\caption{Consistency and symmetry of the operating vibration
patterns. Each row is one instrument, in one regime, under one supply
condition, at one harmonic order $h$ ($h=1,2,3$ for H1, H2, H3). Within a
row the patterns of $n$ fingerings are each compared with that of the
reference fingering A4 by Eq.~(\ref{eq:mac}): MAC is the median of those
$n$ values and MAC$_{\min}$ the lowest single one. $E_a/E$ is the median
antisymmetric energy fraction, Eq.~(\ref{eq:anti_fraction}), which is 0
for pure bending and 1 for pure torsion. Two unrelated patterns give a MAC
of about $1/29 = 0.034$.}
\label{tab:ods_mac}
\small
\setlength{\tabcolsep}{4pt}
\begin{tabular}{llcrrrr@{\hspace{2em}}llcrrrr}
\hline\hline
Instrument & Supply & $h$ & $n$ & MAC & MAC$_{\min}$ & $E_a/E$ & Instrument & Supply & $h$ & $n$ & MAC & MAC$_{\min}$ & $E_a/E$ \\
\hline
\multicolumn{7}{l}{\textit{Reed-dominant (attack window)}} &
\multicolumn{7}{l}{\textit{Tube-dominant (steady window)}}\\
F(i) & low & 1 & 8 & 1.00 & 0.99 & 0.000 & F(i) & low & 1 & 9 & 0.99 & 0.82 & 0.001 \\
F(i) & normal & 1 & 8 & 1.00 & 0.99 & 0.000 & F(i) & normal & 1 & 9 & 0.99 & 0.99 & 0.001 \\
F(ii) & low & 1 & 8 & 0.93 & 0.44 & 0.001 & F(ii) & low & 1 & 9 & 0.41 & 0.07 & 0.203 \\
F(ii) & normal & 1 & 8 & 0.87 & 0.65 & 0.002 & F(ii) & normal & 1 & 9 & 0.70 & 0.35 & 0.044 \\
B$\flat$(i) & low & 1 & 7 & 0.82 & 0.39 & 0.025 & B$\flat$(i) & low & 1 & 8 & 0.13 & 0.00 & 0.043 \\
B$\flat$(i) & low & 2 & 7 & 0.47 & 0.06 & 0.175 & B$\flat$(i) & low & 2 & 8 & 0.10 & 0.00 & 0.175 \\
B$\flat$(i) & low & 3 & 7 & 0.24 & 0.05 & 0.132 & B$\flat$(i) & low & 3 & 8 & 0.08 & 0.00 & 0.296 \\
B$\flat$(i) & normal & 1 & 7 & 0.73 & 0.58 & 0.038 & B$\flat$(i) & normal & 1 & 9 & 0.91 & 0.02 & 0.003 \\
B$\flat$(i) & normal & 2 & 3 & 0.11 & 0.06 & 0.473 & B$\flat$(i) & normal & 2 & 9 & 0.12 & 0.00 & 0.142 \\
B$\flat$(ii) & low & 1 & 8 & 0.81 & 0.04 & 0.013 & B$\flat$(i) & normal & 3 & 9 & 0.25 & 0.01 & 0.338 \\
B$\flat$(ii) & low & 2 & 7 & 0.73 & 0.60 & 0.057 & B$\flat$(ii) & low & 1 & 9 & 0.08 & 0.01 & 0.349 \\
B$\flat$(ii) & low & 3 & 7 & 0.36 & 0.03 & 0.100 & B$\flat$(ii) & low & 2 & 9 & 0.08 & 0.00 & 0.410 \\
B$\flat$(ii) & normal & 1 & 7 & 0.58 & 0.46 & 0.073 & B$\flat$(ii) & low & 3 & 9 & 0.06 & 0.00 & 0.294 \\
B$\flat$(ii) & normal & 2 & 7 & 0.54 & 0.38 & 0.073 & B$\flat$(ii) & normal & 1 & 9 & 0.47 & 0.08 & 0.081 \\
B$\flat$(ii) & normal & 3 & 4 & 0.36 & 0.20 & 0.086 & B$\flat$(ii) & normal & 2 & 9 & 0.25 & 0.01 & 0.169 \\
C(i) & low & 1 & 8 & 0.67 & 0.37 & 0.102 & B$\flat$(ii) & normal & 3 & 9 & 0.10 & 0.00 & 0.261 \\
C(i) & low & 2 & 5 & 0.48 & 0.34 & 0.143 & C(i) & low & 1 & 9 & 0.22 & 0.00 & 0.243 \\
C(i) & low & 3 & 4 & 0.27 & 0.16 & 0.177 & C(i) & low & 2 & 9 & 0.05 & 0.00 & 0.393 \\
C(i) & normal & 1 & 7 & 0.73 & 0.32 & 0.029 & C(i) & low & 3 & 9 & 0.06 & 0.00 & 0.469 \\
C(i) & normal & 2 & 4 & 0.74 & 0.64 & 0.030 & C(i) & normal & 1 & 9 & 0.61 & 0.00 & 0.076 \\
C(i) & normal & 3 & 3 & 0.68 & 0.49 & 0.077 & C(i) & normal & 2 & 9 & 0.13 & 0.02 & 0.330 \\
C(ii) & low & 1 & 8 & 0.97 & 0.11 & 0.005 & C(i) & normal & 3 & 9 & 0.05 & 0.00 & 0.373 \\
C(ii) & low & 2 & 7 & 0.75 & 0.42 & 0.007 & C(ii) & low & 1 & 5 & 0.21 & 0.01 & 0.057 \\
C(ii) & low & 3 & 7 & 0.21 & 0.01 & 0.103 & C(ii) & low & 2 & 5 & 0.05 & 0.00 & 0.477 \\
C(ii) & normal & 1 & 8 & 0.80 & 0.22 & 0.031 & C(ii) & low & 3 & 4 & 0.01 & 0.00 & 0.241 \\
C(ii) & normal & 2 & 8 & 0.76 & 0.06 & 0.025 & C(ii) & normal & 1 & 9 & 0.37 & 0.00 & 0.051 \\
C(ii) & normal & 3 & 7 & 0.38 & 0.14 & 0.093 & C(ii) & normal & 2 & 9 & 0.11 & 0.00 & 0.293 \\
&&&&&& & C(ii) & normal & 3 & 9 & 0.08 & 0.00 & 0.147 \\
\hline\hline
\end{tabular}
\end{table*}
 
\begin{figure*}[!t]
    \centering
    \includegraphics[width=\textwidth]{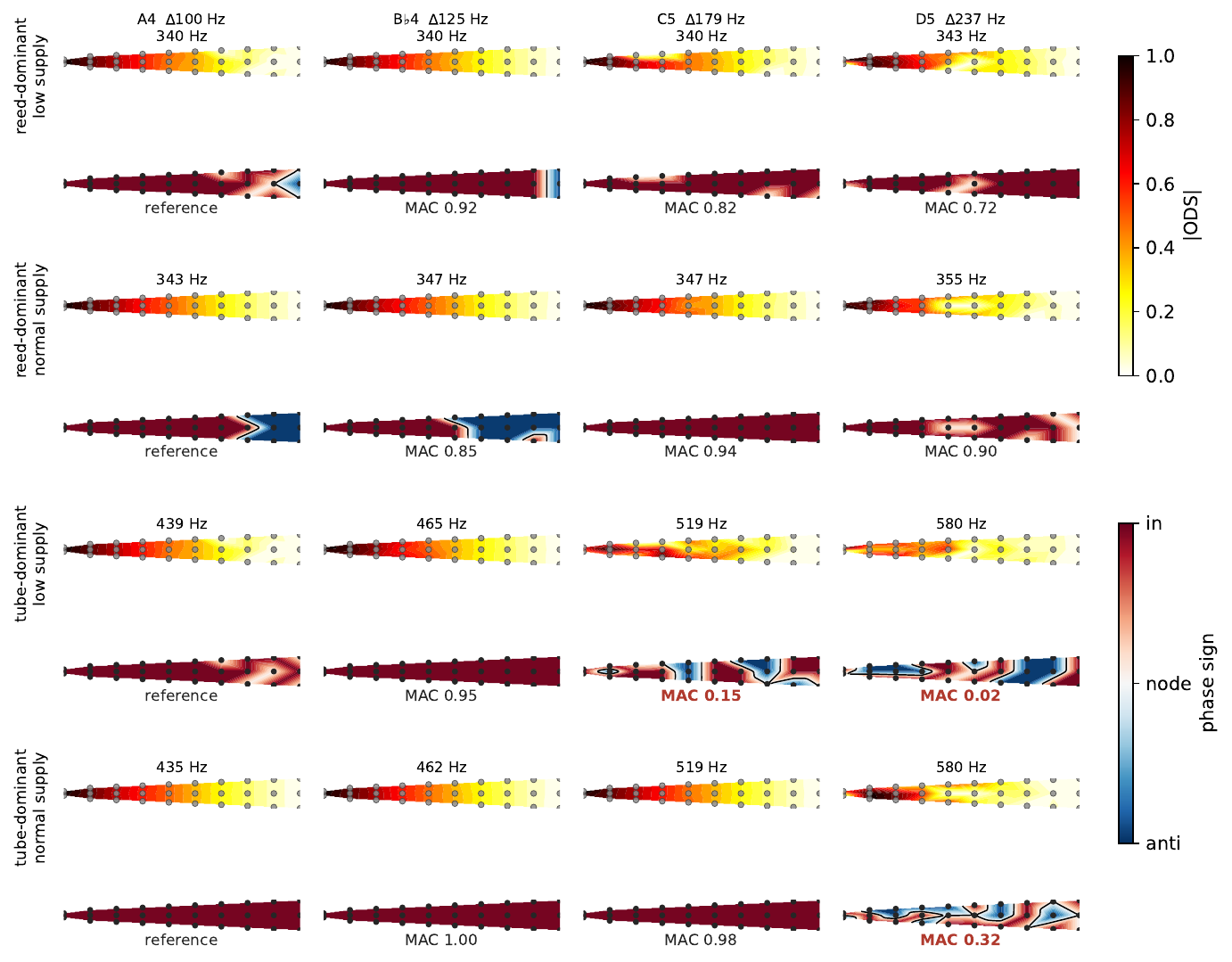}
    \caption{Operating vibration patterns at the fundamental, B$\flat$(i),
    for four fingerings spanning the detuning range. Blocks from top to
    bottom: reed-dominant segment under low and under normal supply,
    tube-dominant segment under low and under normal supply. Within each
    block the upper strip is the pattern normalised to its own maximum and
    the lower strip the sign of the phase, both interpolated between the
    29 scan points (grey circles); the black line is the zero contour of
    the sign field. All patterns are referenced to the tip column, so red
    and blue are comparable between panels. Below each sign strip is the
    MAC against the A4 fingering, the reference throughout; values below
    0.5 are in red. The lowest fingering is omitted, the tube resonance
    there coinciding with the reed frequency.}
    \label{fig:FIG10}
\end{figure*}

\textbf{Tube-dominant state.} Applying the same analysis to the steady-state window gives the opposite picture (lower two groups in Fig.~10). Under normal air supply, the fundamental vibration patterns are highly consistent across fingerings (0.98 and 1.00 against A4 for B$\flat$(i), with median MAC values of 0.47--0.995 across the six instruments). Under low air supply, the C5 and D5 fingerings fall to 0.15 and 0.02 against the same reference, and multiple nodal lines that shift with fingering appear in the sign maps. For five of the six instruments, the median MAC falls to 0.08--0.41 under low air supply. The only exception is F(i), which remains at 0.991 under low pressure. This exception follows the pressure data: the measured reed-side pressure of F(i) under the low-pressure condition is 1.10--1.61~kPa, the highest among all instruments, whereas the corresponding range for B$\flat$(i) is 0.96--1.40~kPa. The instability of the vibration pattern is therefore tied to proximity to the oscillation threshold rather than to the nominal level of the air-supply condition.

The vibration-pattern consistency of both regimes decreases with harmonic order: for the reed-dominant state, from 0.95 at H1 to 0.76 at H2 and 0.42 at H3; for the tube-dominant state, from 0.91 to 0.16 and 0.12. The spatial distribution of the fundamental is robust, whereas those of the higher harmonics depend on fingering and condition.

Torsion about the longitudinal axis was measured by the antisymmetric energy fraction of Eq.~(\ref{eq:anti_fraction}). It carries an independent signature: torsion makes the centerline a nodal line, so the motion vanishes along the centerline and appears on both sides in antiphase. Cottingham used this signature to identify the first torsional mode in harmonium reeds\citep{Cottingham2013}, and it is used here as a cross-check by taking, column by column, the amplitude ratio between the centerline and the two sides. The reed-dominant state is essentially pure bending: the median fraction is 0.021 (interquartile range 0.003--0.093) and does not change with the air-supply condition. The fraction rises with harmonic order (0.062 for H2 and 0.115 for H3), is higher in the tube-dominant state (0.050 for H1, 0.194 for H2, and 0.323 for H3), and is largest under low air supply (median 0.222). The centerline-to-side amplitude ratio is 0.98 to 1.08 in the onset window and stays close to 1 in the tube-dominant segment, which agrees with the antisymmetric energy fraction. The torsional component is therefore not an intrinsic property of the reed, it is excited when air-column loading dominates and the driving is weaker.

The reeds of Western reed organs usually carry an intentional curvature and a slight twist at the free end of the tongue, which makers consider to improve the tone quality and shorten the onset; experiments show that this shape does excite the first torsional mode and shorten the onset time\citep{Cottingham2013}. Asian free reeds are cut from a single piece of material and are approximately symmetric with respect to the airflow direction, without such built-in twist. The torsional component found here is near zero at the reed-dominant fundamental, which follows the construction of the reed.

\textbf{Vibration pattern and phase reference after decoupling.} The reappearing oscillation segment after the loss of locking described in Sec.~IV.A requires a different phase reference. This segment is preceded by a silent interval of approximately 100~ms, during which the amplitude falls to about 47~dB below the steady-state level. The re-onset that follows is triggered by a noise seed whose phase is not repeatable between triggering events. Because the velocity is measured point by point, the velocity phases at different points are not comparable: the cross-fingering MAC computed from the velocity phase is 0.079, which is the level expected of unrelated 29-dimensional complex vectors.

At each scan point the setup records a microphone signal synchronously with the velocity. The unknown seed phase is common to the velocity and the microphone signal at that point, so the transfer function of Eq.~(\ref{eq:transfer}) gives the phase referenced to the sound field of that event, the seed cancelling in the ratio. The microphone does carry the reed fundamental during this segment: the in-band signal-to-noise ratio for the four fingerings ranges from 27.5 to 30.3~dB, the median coherence from 0.81 to 0.88, and all 29 points are usable.

After referencing to the microphone signal (Fig.~11), the cross-fingering MAC rises from 0.079 to 0.999. The phases of the 29 points agree, with standard deviations of 0.57$^\circ$--0.92$^\circ$; the median phase difference between mirror-point pairs is 0.3$^\circ$--0.5$^\circ$, with a maximum of 3.0$^\circ$. Torsion about the longitudinal axis requires mirror pairs to be in antiphase, a difference near 180$^\circ$, far from these values. The corresponding antisymmetric energy fraction is $2\times10^{-5}$ to $9\times10^{-5}$. The amplitude decreases monotonically along the reed from 1.00 at the tip to 0.02 at the root, a tip-to-root ratio of 49--53, with no node over the entire reed. In the onset window of the same decoupling sequence the cross-fingering MAC is 0.885, the same pattern as the 0.999 found after decoupling.

\begin{figure*}[!t]
    \centering
    \includegraphics[width=\textwidth]{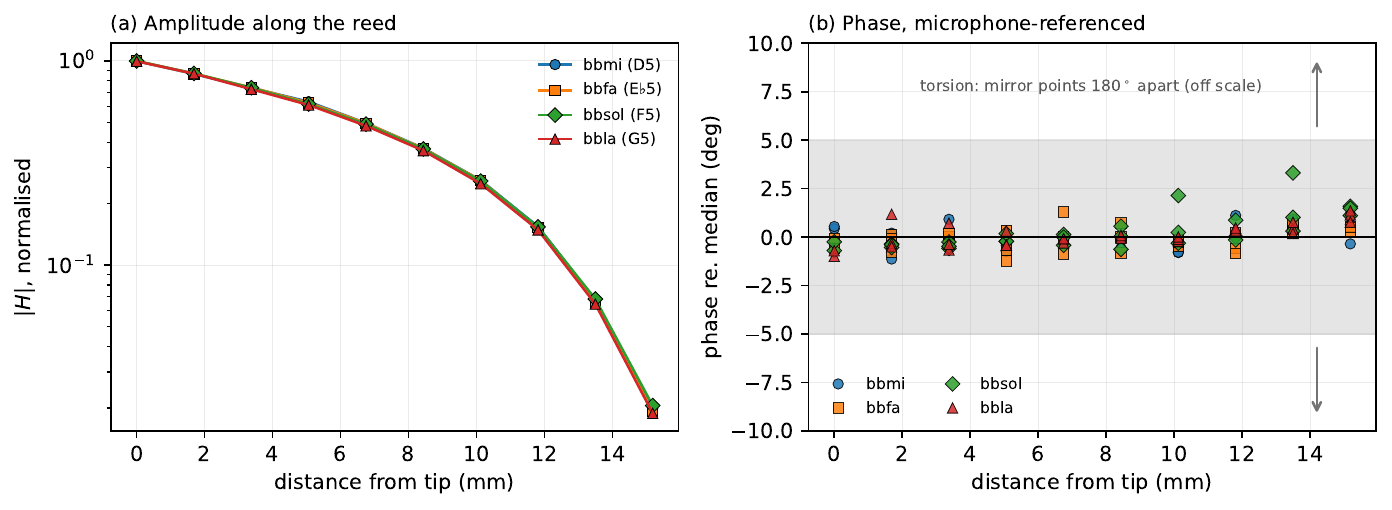}
    \caption{Reed-dominant segment following unlocking, B$\flat$(i), four
    fingerings of the decoupling series, reconstructed with the
    microphone (Ref3) as phase reference, Eq.~(\ref{eq:transfer}).
    (a) $|H_i|$ along the reed, averaged across the three scan lines and
    normalised to its maximum. (b) Phase of $H_i$ at every scan point
    relative to its median. Arrows mark the $180^\circ$ separation
    between mirror partners that torsion about the long axis would
    require.}
    \label{fig:FIG11}
\end{figure*}

In summary, the reed-dominant state is consistent in the frequency, temporal, and spatial dimensions. It is pure first-order bending across all six instruments and all fingerings, and it recovers after decoupling with the same frequency and the same pattern; the spatial distribution of the tube-dominant state is set by air-column loading and becomes unstable near the oscillation threshold.

\section{\label{sec:5}Discussion}
\subsection{Detuning as an Organizing Parameter}

The four sets of observations in this study are organized by the same quantity: the detuning between the reed and the air column, $\Delta = f_\mathrm{tube} - f_\mathrm{reed}$.

The reed-dominant component is the reed itself, on three counts. In the steady-state spectrum it stands in no integer ratio with $f_\mathrm{tube}$, so it is not a harmonic. The oscillation that reappears after the loss of locking returns to it within 0.1~Hz (Sec.~IV.A). And the spatial patterns of the two segments are the same first-order bending pattern, with a microphone-referenced MAC of 0.999 (Sec.~IV.D). Frequency, time, and space agree.

The quantity $\Delta$ then organizes the rest. At its lower limit, $\Delta \to 0$, the two components are no longer distinguishable and the system enters frequency capture\citep{Bader2013}; the amplitude modulation reported here occurs only for the captured fingering (Sec.~IV.C). In time, the silent interval after the loss of locking and the durations of the two regimes vary monotonically with $\Delta$ (Fig.~6). In space, under the low air-supply condition, the tube-dominant vibration patterns cease to repeat across fingerings once $\Delta$ exceeds about 150~Hz (Fig.~10). $\Delta$ therefore ties the frequency, time, and space observations together.

Cottingham's measurements of the \emph{bawu} provide a direct precedent: opening tone holes shortens the effective tube length, which raises the pitch and also changes the timbre, since reed--tube coupling carries the frequency away from the reed's natural frequency and the tone comes to be set by the tube resonance\citep{Cottingham2011}. The $\Delta$ staircase measured here puts a number on that observation: within one instrument $\Delta$ rises from about 60~Hz at the lowest sounding fingering to 419~Hz at the highest, while the reed-dominant component stays present as a separate spectral peak throughout. Cottingham also noted that the playable range of reed--tube combinations in instruments such as the \emph{bawu} is usually less than an octave; the nine fingerings of each of the six instruments here span about an octave.

\subsection{Pressure Dependence and the Outward-Striking Configuration}

Two observations together fix the operating configuration of the reed.

First, the coupled sounding frequency always lies above the reed-dominant frequency: $\Delta$ is positive in all 371 valid observations, running from 38~Hz to 523~Hz, while the ratio $R$ runs from 1.15 to 2.49. Under Fletcher's onset condition, $\omega > \omega_r$ belongs to the $(+,-)$ outward-striking configuration, the $(-,+)$ inward-striking configuration requiring the opposite. This matches the predictions and measurements of Cottingham and Dieckman for Asian reed--tube combinations, where the sounding frequency lies above both the natural frequency of the reed and the tube resonance\citep{Cottingham2009,Dieckman2006}. Hikichi \emph{et al.} reached the same configuration for the \emph{sho} from a time-domain model checked against artificially blown measurements\citep{Hikichi2003}.

Second, the reed-dominant frequency rises monotonically with the pressure on the reed side, by about 160~cents per kPa (Sec.~IV.B). In the outward-striking configuration a larger pressure difference raises both the mean opening and the aerodynamic stiffness term\citep{StHilaire1971}, carrying the frequency upward, so the positive sign agrees with the first observation.

The $f_\mathrm{reed}$ measured here comes from the onset window before coupling is established. It is the self-sustained oscillation frequency of the reed before the air column takes over, not the mechanical resonance frequency $f_r$ of the unblown reed. The two run in the same direction but are not the same quantity; $f_r$ would come from the free decay of the unblown reed under impulse excitation, which would allow the onset criterion\citep[Eq.~(13.8)]{Fletcher1998} to be applied directly.

\subsection{Vibration-Pattern Stability Depends on the Distance from the Oscillation Threshold}

Within one instrument, the tube-dominant pattern is more consistent under normal supply than under low supply. This holds in all five instruments that yielded usable scans under both conditions, C(ii) low having passed none (five of five, sign test $p = 0.031$). The relationship does not carry across instruments: absolute reed-side pressure is not associated with MAC (Spearman $\rho = 0.26$, $p = 0.45$)---F(i) already reaches 0.991 at 1.42~kPa, whereas C(ii) reaches only 0.366 at 1.81~kPa. What decides whether the vibration pattern is stable is therefore not the absolute pressure but how far the oscillation state sits from the onset threshold of that particular instrument, which varies with reed geometry and assembly.

\subsection{Breakdown of Locking}

Cottingham noted that free-reed instruments start slowly, the sound building up gradually after oscillation begins, and that the effect grows for longer reeds and lower pitches\citep{Cottingham2013}. For B$\flat$(i), the coupled state is established about 125~ms after the onset of reed oscillation (Sec.~IV.A).

The collapse of the coupled state is not triggered by too small a pressure difference. It occurs at 5.1 times the reed oscillation threshold and at 70\% of the peak pressure, and during the silence that follows the pressure stays above four times the threshold while the reed remains below the noise floor for about 35 cycles (Sec.~IV.A). Were the reed motion set by the pressure difference alone, it would recover at once. The collapse and the silence belong to the coupled system.

Locking is established at $p = 1.50$ on the rising branch and breaks down at $p = 1.08$ on the falling branch. The inequality fits bistability, but the pressure sweep here is not quasistatic and the transition points carry rate-dependent delays; one sweep cannot separate true bistability from dynamic hysteresis under a supercritical bifurcation. Separating them calls for repeated upward and downward ramps at controlled rates, together with a bistability test holding the pressure constant inside the window. The amplitude modulation of Sec.~IV.C---modulation of the orbit radius of a single limit cycle, with the pressure envelope nearly constant throughout---likewise calls for an independent measurement of the transverse stability of the limit cycle near the threshold. Both lie outside the present study and will be reported elsewhere.

\subsection{Method: Phase Reference after Restart of Self-Sustained Oscillation}

Point-by-point laser vibrometry gives a meaningful complex vibration pattern as long as the oscillation phase is locked to the trigger. The onset segment and the frequency-locked steady-state segment satisfy this; the reappearing segment after decoupling does not. The amplitude first falls to about 47~dB below the steady-state level, and the onset that follows is triggered by a noise seed whose phase is not repeatable between triggering events. The cross-fingering MAC computed from the velocity phase therefore drops to 0.079, the level expected of random complex vectors.

The microphone signal recorded synchronously with the velocity supplies a reference: the unknown seed phase is common to the two signals at that point and cancels in the transfer function of Eq.~(\ref{eq:transfer}). After referencing, the MAC rises to 0.999 and the phase over the whole reed is consistent to within 0.6$^\circ$--0.9$^\circ$.

\subsection{Limitations}

The vibration patterns reported here are the spatial distributions of a nonlinear limit cycle at its fundamental and harmonics, not structural eigenmodes. The measured frequencies stand in strict integer ratios and do not match the modal frequency ratios of a cantilever reed, so the term ``mode'' does not apply. The measurements were made under simulated playing conditions, without lip constraint or vocal-tract coupling, and the difference from actual playing has not been quantified. The scan grid of three longitudinal lines with ten stations each resolves the presence or absence of a first-order torsional component and no finer transverse detail. Cottingham further detected a transverse mode in accordion reeds whose motion lies in the plane of the unexcited reed\citep{Cottingham2013}; the laser vibrometer reads the out-of-plane velocity component, so such in-plane motion does not appear in these measurements. Amplitude modulation was seen for one fingering of one instrument, and the decoupling sequence covered four fingerings of one instrument; how far either generalizes is still open.

\section{Conclusions}

The vibration of the reeds of six \emph{hulusi} was measured point by point by laser vibrometry under simulated playing conditions.

\begin{enumerate}

\item
The velocity spectra of steady-state oscillations contain two resolvable components: the tube-dominant frequency, locked to the air-column resonance and carrying a complete harmonic series, and the reed-dominant frequency, which is not in an integer ratio with it. Their difference $\Delta$ runs from 38~Hz to 523~Hz across the dataset and the ratio from 1.15 to 2.49. For fingerings with $\Delta \to 0$ the two components are no longer separable and the system enters frequency capture.

\item
The reed-dominant frequency rises monotonically with the pressure on the reed side, by about 160~cents per kPa, in the same direction in all eleven instrument--condition units (sign test, $p = 5\times10^{-4}$). The pressure effect holds once the air-column resonance is controlled for ($\partial f/\partial p = +28.0 \pm 10.3$~Hz/kPa, $p = 0.008$), whereas air-column loading contributes nothing of its own once pressure is controlled for. The positive sign belongs to the outward-striking $(+,-)$ configuration. That configuration follows from the construction---tongue and frame cut from one piece, with a permanent set holding the free end clear of the plate (Sec.~III~D)---and is supported independently by $\Delta > 0$ in every observation.

\item
After the air supply is switched off, the coupled state collapses within a narrow interval and, after a silence of about 35 cycles, the reed resumes oscillation at its own frequency, 0.1~Hz from the onset segment. The collapse occurs at 5.1 times the reed oscillation threshold, so too small a pressure difference does not account for it. The length of the silence and the durations of the two regimes vary monotonically with $\Delta$.

\item
The fundamental vibration pattern of the reed-dominant state is pure first-order bending, the same across all six instruments and all fingerings (MAC 0.58--0.998), with an antisymmetric energy fraction of 0.021. The tube-dominant vibration pattern repeats across fingerings only well away from the oscillation threshold; near the threshold its spatial distribution and its torsional component become unstable together.

\item
The oscillation that reappears after decoupling carries no detectable torsion: the microphone-referenced complex vibration pattern gives a whole-reed phase consistency of 0.6$^\circ$--0.9$^\circ$, a phase difference of 0.3$^\circ$--0.5$^\circ$ between mirror-point pairs, and an antisymmetric energy fraction of order $10^{-5}$. This rests on a phase reference recorded synchronously with the motion; without one, the phase obtained point by point after the restart of self-sustained oscillation falls inside the random null.

\end{enumerate}

These results place the hulusi between two kinds of behavior. At normal playing pressure the tube sets the pitch and the finger holes change it, as in a flute. At low pressure the reed sets the pitch, the finger holes change the timbre instead, and the vibration pattern of the tube-dominant regime no longer carries from one fingering to the next. The instrument is played at normal pressure, but the pressure rises through the low range at every attack, so the reed-dominant regime measured here is the state the sound passes through before it locks. Its frequency, its duration and its vibration pattern are all set by $\Delta$, which gives the attack of the hulusi a dependence on fingering that a flute does not have and places the instrument with the \emph{sheng} and the \emph{sho} rather than with the beating-reed instruments.

\section*{Author Declarations}
\subsection*{Conflict of Interest}
The authors have no conflicts to disclose.

\subsection*{Data Availability}

The data that support the findings of this study are available from the corresponding author upon reasonable request.

\vspace{1cm}
\textbf{Literature}

\bibliography{citation-1}

\end{document}